\documentclass[times,twocolumn,final,authoryear]{elsarticle}

\usepackage{jasr}
\usepackage{framed,multirow}
\usepackage{amssymb}
\usepackage{amsfonts} 
\usepackage{amsmath}
\usepackage{latexsym}
\usepackage{url}
\usepackage{xcolor}
\usepackage[citebordercolor=white]{hyperref}
\usepackage{hyperref}
\hypersetup{colorlinks,
citecolor=blue
}
\usepackage{graphicx}
\usepackage{subcaption}
\usepackage{hyperref}
\usepackage{hypcap}
\usepackage{wrapfig,booktabs}
\usepackage[export]{adjustbox}
\usepackage{needspace}
\usepackage{bibentry}
\usepackage{comment}

\usepackage[switch]{lineno}
\journal{Advances in Space Research}
\graphicspath{{figures/}}

\definecolor{newcolor}{rgb}{.8,.349,.1}
\definecolor{ao(english)}{rgb}{0.0, 0.5, 0.0} 

\newcommand{\BE}{\begin{equation}}
\newcommand{\EE}{\end{equation}}
\newcommand{\BA}{\begin{align}}
\newcommand{\EA}{\end{align}}

\newcommand{\ie}{i.e.}

\newcommand{\degree}{\ensuremath{^\circ}}

\newcommand{\angstrom}{\textup{\AA}}

\begin{document}
\verso{Maharana \textit{et al}}

\begin{frontmatter}

\title{Implementation and validation of the FRi3D flux rope model in EUHFORIA} 

\author[1,5]{Anwesha \snm{Maharana}\corref{cor1}} 
\author[2]{Alexey \snm{Isavnin}}
\author[3,4]{Camilla \snm{Scolini}}
\author[1]{Nicolas \snm{Wijsen}}
\author[5]{Luciano \snm{Rodriguez}}
\author[5,6]{Marilena \snm{Mierla}}
\author[1,5]{Jasmina \snm{Magdaleni\'c}}
\author[1,7]{Stefaan \snm{Poedts}}

\address[1]{CmPA/Department of Mathematics, KU Leuven, Celestijnenlaan 200 B, 3001 Leuven, Belgium}
\address[2] {Rays of Space, Finland-Belgium}
\address[3]{Institute for the Study of Earth, Oceans, and Space, University of New Hampshire, Durham, NH, 03824 USA}
\address[4]{CPAESS, University Corporation for Atmospheric Research, Boulder, CO, 80301 USA}
\address[5]{Solar–Terrestrial Centre of Excellence—SIDC, Royal Observatory of Belgium, 1180 Brussels, Belgium}
\address[6]{Institute of Geodynamics of the Romanian Academy, Bucharest, Romania}
\address[7]{Institute of Physics, University of Maria Curie-Sk{\l}odowska, Pl. M. Curie-Sk{\l}odowskiej 5, 20-031 Lublin, Poland}

\received{27 November 2021}
\finalform{2021}
\accepted{2021}
\availableonline{2021}
\communicated{}

\begin{abstract}
The "Flux Rope in 3D" \citep[FRi3D,][]{Isavnin2016}, a coronal mass ejection (CME) model with global three-dimensional (3D) geometry, has been implemented in the space weather forecasting tool EUHFORIA \citep[][]{Pomoell2018}. 
By incorporating this advanced flux rope model in EUHFORIA, we aim to improve the modelling of CME flank encounters and, most importantly, the magnetic field predictions at Earth. After using synthetic events to showcase FRi3D's capabilities of modelling CME flanks, we optimize the model to run robust simulations of real events and test its predictive capabilities. 
We perform observation-based modelling of the halo CME event that erupted on 12 July 2012. 
The geometrical input parameters are constrained using the forward modelling tool included in FRi3D with additional flux rope geometry flexibilities as compared to the pre-existing models. 
The magnetic field input parameters are derived using the differential evolution algorithm to fit FRi3D parameters to the in-situ data at 1 AU. 
An observation-based approach to constrain the density of CMEs is adopted, in order to achieve a better estimation of mass corresponding to the FRi3D geometry. 
The CME is evolved in EUHFORIA's heliospheric domain and a comparison of FRi3D’s predictive performance with the previously implemented spheromak CME in EUHFORIA is presented. 
For this event, FRi3D improves the modelling of the total magnetic field magnitude and $B_z$ at Earth by ${\sim}30\%$ and ${\sim}70\%$, respectively. Moreover, we compute the expected geoeffectiveness of the storm at Earth using an empirical $Dst$ model and find that the FRi3D model improves the predictions of minimum $Dst$ by ${\sim}20\%$ as compared to the spheromak {\color{black}{CME model}}. 
Finally, we discuss the limitations of the current implementation of FRi3D in EUHFORIA and propose possible improvements. 
\end{abstract}

\begin{keyword}
\KWD Sun\sep Coronal mass ejections\sep Global flux rope\sep Magnetohydrodynamics\sep Heliosphere\sep Geoeffectiveness
\end{keyword}

\end{frontmatter}

\section{Introduction}
\label{sec:Introduction}
Coronal Mass Ejections (CMEs) are giant clouds of magnetized plasma erupting from the Sun. They are responsible for the majority of strong geomagnetic storms \citep[see e.g.][]{zhang:2007}. CMEs typically emerge from the unstable twisting and shearing of the magnetic field structures in the active regions of the Sun and manifest a magnetic flux rope structure \citep[][]{Priest2002,Schmieder2006,Jiang2021,Vemareddy2021}. After the eruption, CMEs propagate in the interplanetary medium and we refer to them as Interplanetary CMEs \citep[ICMEs,][]{Gosling1980, Zurbuchen2006}. While propagating through the corona and the interplanetary space, they undergo changes upon interacting with e.g., the solar wind plasma \color{black}{\citep{Winslow2021} or} another ICME \citep{Kilpua2019}. Although most of the evolution happens in the lower corona, CMEs are known to exhibit rotations and deflections at larger heliocentric distances too \citep[][]{Wang2004a,Vourlidas2011}. Such effects can play a crucial role in determining the exact propagation direction of the Earth-directed CMEs and the specifics of their magnetic field configurations in terms of strength and orientation, which influence the level of their geomagnetic impact \citep{Isavnin2014,Winslow2016,Winslow2021}. As the Earth-directed CMEs pose a potential threat to technology, economy, and life on Earth, it is important to accurately forecast their arrival time and geoeffectiveness. 

There are different approaches on how to track the CME evolution from Sun to Earth using observations and models \citep{Isavnin2013,Kay2013,Kilpua2013,rodriguez2008}. These previous works consider only a limited subset of the CME properties and heavily depend on observations to determine the CME structure. An alternative approach is to use three-dimensional (3D) CME models in magnetohydrodynamic (MHD) simulations of the solar wind such as e.g., ENLIL \citep{Odstrcil2003}, EUHFORIA \citep{Pomoell2018}, SUSANOO-CME \citep{Shiota2016}, MS-FLUKSS \citep[][]{Singh2018}. 
CMEs are inserted in these models at 0.1~au (where the solar wind is assumed to be super-Alfv\'{e}nic and super-fast) and are then self-consistently evolved by solving the MHD equations. 

The initial CME parameters that need to be determined and extrapolated to $0.1$~au rely heavily on CME reconstruction methods. 
Multiple viewpoint reconstruction techniques using white light coronagraph images include the Graduated Cylindrical Shell model \citep[GCS, ][]{Thernisien2011} and StereoCAT \citep[][]{Mays2015}. These are the most widely used tools to obtain CME parameters for initializing models like EUHFORIA and ENLIL. Although the forward modelling tool of GCS performs quite well, it does not take into consideration the deformations of the flux rope due to rotational skew, radial expansion or pancaking, and front flattening. However, it is crucial to determine the boundary conditions as accurately as possible for driving the heliospheric propagation of the CMEs in the MHD models.

\color{black}{When observed in situ, many ICMEs do not show a clear magnetic flux rope structure (e.g. absence of smooth magnetic field rotation with low plasma beta), such events are more difficult to model \citep{Rodriguez2004}. 
Approximately 70\% of ICMEs during solar minima contain magnetic clouds while during the solar maxima this number is significantly lower amounting to about 20\% 
\citep{Richardson2004}.}

The study by \citet{Marubashi2007} points out that a large number of spacecraft encounters (at Earth) with magnetic clouds occur at the CME flanks. This highlights the requirement of advanced extended geometry, like that of a torus rather than a cylinder, to interpret the curved portion of the magnetic cloud in the cases where the spacecraft traversed the flank of the magnetic cloud loop.

In \citet{Isavnin2016}, a novel flux rope CME model FRi3D with a 3D magnetic field configuration and a global CME geometry has been introduced to better reproduce ICME observations in the heliosphere. In this study, we implement FRi3D in EUropean Heliospheric FOrecasting Information Asset \citep[EUHFORIA, ][]{Pomoell2018} and compare its performance with the pre-existing CME models. 
The aim is to assess the performance of the FRi3D model in predicting the ICME parameters at Earth and compare it with the existing flux rope CME model in EUHFORIA, \ie, the spheromak model \color{black}{as described in} \citet{Verbeke2019}. The spheromak model is an upgrade over the unmagnetized cone model as it can predict the magnetic field of the ICME at Earth by virtue of its linear force-free internal magnetic field. However, like any other model, spheromak also shows certain drawbacks highlighted in \citet{Scolini2019}. Due to its compact spherical shape and lack of CME legs, spheromak fails to model the events where ICMEs impact Earth with their flanks. In addition, with the spheromak model the strengths of the magnetic field components are underestimated, which makes the prediction of the geoeffectiveness difficult as it strongly depends on the $B_z$ component of the flux rope \citep[][]{Kane2010}. Due to the presence of legs in the FRi3D flux rope, the model manages to capture the essential signatures of the CME flanks and their helio-effectiveness in the interplanetary medium.     

In the next section, we provide an introduction to the FRi3D flux rope and the EUHFORIA model. In Section~\ref{sec:Implementation}, the implementation of the FRi3D flux rope model in EUHFORIA is described.
Section~\ref{sec:Case_synthetic} shows the results of the validation study of the CME model using a synthetic CME simulation. We extend the validation study to an observed CME event which is detailed in Section~\ref{sec:Case_real}. Finally, in Section~\ref{sec:Discussion} we summarize and discuss the results of this study and present future improvements and development plans.  

\section{Models}
\label{sec:Models}
\subsection{The FRi3D model}

\begin{figure}[htb!]
    \centering
    \includegraphics[trim={2cm 0 0 0},clip,width=10cm,height=9.5cm]{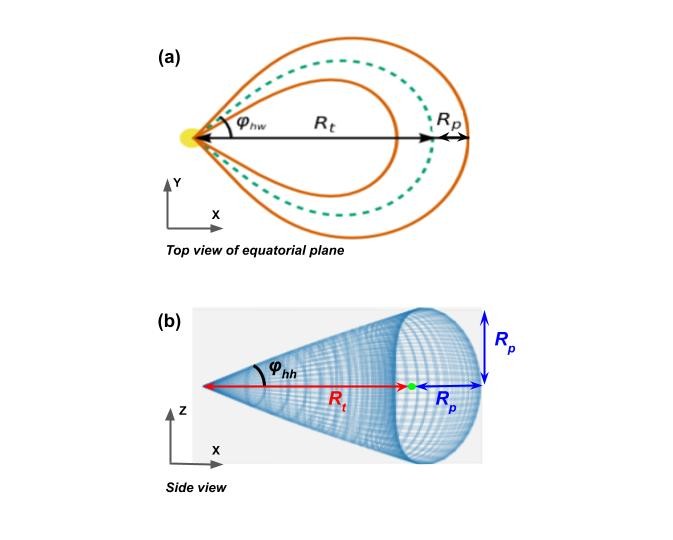}
    \caption{FRi3D flux rope geometry. (a) Figure adapted from \citet{Isavnin2016}, shows the view of the FRi3D flux rope on the equatorial plane{\color{black}{. $\phi_{hw}$ is measured from the axis (green dashed line)}}; (b) The side view of the geometrical parameters as provided in Table~\ref{tab:fri3d_params_description}. {\color{black}{The horizontal red and blue arrows are drawn schematically and their meeting point (green dot) is at the axis of the CME.}}
    }
    \label{fig:geo_params}
\end{figure}

\begin{figure}[htb!]
    \centering
    \includegraphics[trim={0.25cm 0 0 0},clip,width=9cm, height=5.6cm]{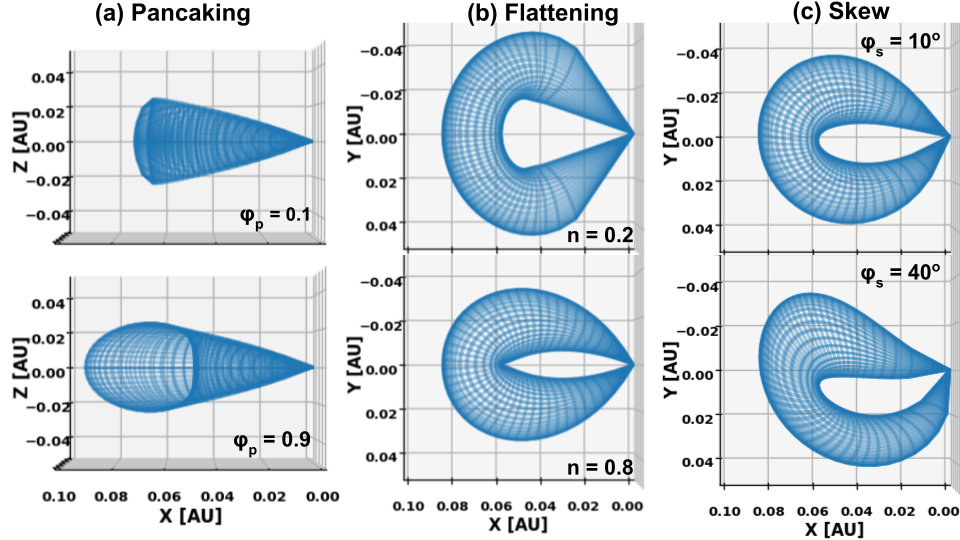}
    \caption{{\color{black}{Variation of the FRi3D model shell in 3D due to different pancaking, flattening and skew parameters.}} The figure shows (a) pancaking [$\varphi_p = 0.1 \text{ (top)}, 0.9 \text{ (bottom)}$], (b) front flattening [$n = 0.2 \text{ (top)}, 0.8 \text{ (bottom)}$] and (c) skew [$\varphi_s = 10\degree \text{ (top)}, 40\degree \text{ (bottom)}$] deformations. Each pair of figures (top and bottom) {\color{black}{represents}} extreme manifestations of the deformation {\color{black}{parameter reported in each plot}}.} 
    \label{fig:shell_deformation}
\end{figure}

The "Flux Rope in 3D" \citep[FRi3D,][]{Isavnin2016} is a fully analytical 3D model of a flux rope CMEs that is capable of reproducing the global geometrical shape of a CME as well as all of its major deformations. 
It is constructed in three steps. First, the global axis of the model is derived. It is estimated as a shape of a magnetized slingshot in a radial non-magnetized flow. The shape of the slingshot in such a setup is defined by the balance of magnetic tension of the slingshot and the hydrodynamic forces of the flow. Second, FRi3D consists of a cylindrical shell with changing curved non-circular axis and variable cylinder radius, hence making it more flexible than a torus. The shell around the global axis is added resulting in a smooth continuous croissant-like 3D shape, as illustrated in Figure~\ref{fig:geo_params}. The parameters of the axis and the shell mimic geometrical deformations such as pancaking, flattening, and rotational skew, as shown in Figure~\ref{fig:shell_deformation}. {\color{black}{Pancaking is deformation that describes the tendency of the flux rope frontal part to expand in poloidal and azimuthal directions (due to radial propagation) while experiencing relatively smaller growth in radial thickness. Due to this deformation, the frontal part of a CME will start to resemble a pancake. The flattening parameter accounts for the flattening of the front part of the flux rope, which can be observed when the faster flux rope propagates through the significantly slower solar wind. Skew refers to the deformation of the flux rope connected to the Sun, due to the accumulated solar rotation.}} Having a fully analytic shell, FRi3D can be used to model geometrical transformations like e.g., rotation, deflection, and expansion. 


Finally, this highly flexible shell is populated with magnetic field lines. The magnetic field strength is estimated using the force-free magnetic field distribution in cylindrical geometry, with helical and twisted field lines characterized by the Lundquist model \citep{Lundquist1950}. Magnetic field lines are subjected to the same pancaking and skew deformations that the shell undergoes, to tailor the field as per the shape of the extended and deformed FRi3D shell. {\color{black}{The flux rope is not completely force-free as a result of introduced deformations. However, the observed CMEs are also not completely force-free either \citep{Martin2020}. Mathematically, inserting a non-force-free model into an MHD simulation is challenging but for certain combinations of parameters, it gets balanced during the propagation and does not show excessive size increase.}} 

 The FRi3D model is available as an open-source package (\url{https://pypi.org/project/ai.fri3d/}) which enables its further development and usage as a tool. The model can employ both remote and in-situ observations (at any heliocentric distance) to estimate the global shape and the internal magnetic field structure of the CME \citep{Isavnin2016}. The model can perform 3D reconstruction of CMEs using white light images from multiple vantage points, and it complements the GCS model by adding deformation information within $0.1$~au. The in-situ observations fitting algorithm of FRi3D enables constraining of the magnetic field parameters during the spacecraft trajectory through the CME, for any spacecraft-CME encounter geometry. These FRi3D tools will be used for constraining initial parameters to model the event discussed in Section~\ref{sec:Case_real}.
 
\subsection{The EUHFORIA model}
The EUropean Heliospheric FORecasting Information Asset \citep[EUHFORIA,][]{Pomoell2018} is a physics-based solar wind and CME propagation model designed for space weather studies and forecasting purposes. The model has two parts, the coronal domain extending up to the inner heliospheric boundary of $0.1\;$au and the heliospheric domain covering distances beyond $0.1\;$au. The heliospheric inner boundary is assumed to be beyond the Alfv\'{e}n surface, where the solar wind becomes super-Alfv\'{e}nic and super-fast \citep{Parker1965,Weber1967}. 
The coronal model used in EUHFORIA is the {\color{black}{semi-}}empirical Wang-Sheeley-Arge model \citep[WSA;][]{Arge2003} which employs the synoptic maps of the line-of-sight photospheric magnetic field to compute the plasma parameters at $0.1\;$au. 
The WSA model implemented in EUHFORIA extrapolates the magnetic field radially through the Potential Field Source Surface (PFSS) model up to $2.6$ solar radii (hereafter $R_\odot$) in the lower corona, and via the Schatten Current Sheet (SCS) model in the upper corona extending from $2.3 R_\odot$ to $21.5 R_\odot$. In the WSA model, the solar wind speed is an empirical function of the flux tube areal expansion factor and the distance from the foot of open field lines to the coronal hole boundary. A detailed analysis of the coronal model parameters can be found in \citet{Asvestari2019}.
The plasma parameters obtained from the coronal module act as the initial conditions to solve the 3D time-dependent ideal MHD equations in Heliocentric Earth Equatorial (HEEQ) coordinate system. 
After obtaining the solar wind in the heliospheric domain, CMEs can be inserted at 0.1~au as a time-dependent boundary condition, after which their evolution, arrival, and impact on Earth can be studied. 

A brief description of the two presently available CME models in EUHFORIA is listed below: 

\begin{itemize}
    \item Cone: In the ``full ice-cream cone model”, the CME is a hydrodynamic cloud of plasma without any intrinsic magnetic field \citep[see, e.g.][]{Xue2005,Gopalswamy2009(a), Na2017}. {\color{black}{A}} spherical blob is self-similarly expanded, conserving the angular width, propagation direction, and speed, before being launched into the heliospheric domain. Although it cannot predict the magnetic field, its simplistic geometry and limited parameters simplify its usage and provide a stable simulation running environment. It is useful for calculating CME arrival times and arrival locations.

    \item Spheromak: In the spheromak model, the CME magnetic field configuration is that of a magnetized linear force-free flux rope in spherical geometry \citep[][]{Chandrasekhar1957,Verbeke2019}. It is launched with a uniform speed, density, and temperature like in the cone model. Its geometry is similar to Cone CME, like a spherical blob. It complements the Cone model by predicting the magnetic field at Earth. The spherical shape was also used to overcome problems with modelling the effect of footpoints of magnetized CMEs while simulating multiple eruptions. The spheromak is completely pushed through the inner boundary without leaving any foot point traces. 
 
\end{itemize}

\renewcommand{\arraystretch}{1.25}
\begin{table*}
\centering
\begin{tabular}{ p{2.8cm} p{11.0cm} p{2.2cm}  }
 \hline
 \multicolumn{3}{c}{\textbf{FRi3D parameters}} \\
 \hline
 Parameter   &  Description & Value range \\ \midrule
 Latitude $\theta_{CME}$    & Polar angle of launch position in spherical HEEQ coordinates [degree] &  [$-60$, $60$]\\ 
 Longitude $\phi_{CME}$   & Azimuthal angle of launch position in spherical HEEQ coordinates [degree] &  [$-180$, $180$]\\
 Half-width $\varphi_{hw}$ & Maximum angular extent in the azimuthal direction from the line joining the origin to the apex of the flux rope [degree] &  [10, 75]\\
 Half-height $\varphi_{hh}$ & Maximum angular extent in the polar direction from the CME plane [degree] &  [$5$, $60$]\\
 Toroidal height $R_t$ & Heliocentric distance to the apex of CME axis [$R_\odot$] &  [$0$, $21.5$]\\
 Flattening $\textit{n}$ & Deformation at the CME front due to high speed launch [unitless] & [$0.2$, $0.8$]\\
 Pancaking $\varphi_p$  & Deformation due to radial expansion [unitless] & [$0.1$, $0.9$]\\
 Skew $\varphi_s$        & Deformation due to solar rotation (set to zero in EUHFORIA implementation) [degree] & [$-50$, $50$]\\ \midrule
 Chirality    & Handedness of the flux rope (In EUHFORIA simulations, value for right-handed and left-handed FRi3D flux ropes are -1 and +1 respectively.) [unitless] & $\pm1$ \\  
 Polarity    & Direction of axial magnetic field of the flux rope (+1 corresponds to east-to-west direction of magnetic field from footpoint to footpoint) [unitless] & $\pm1$ \\
 Tilt        & Angular orientation of CME axis measured from equatorial plane using right-hand rule around the axis with origin at the Sun [degree] & [$-180$, $180$] \\
 Magnetic flux $\phi_B$ & Total magnetic flux of CME [Wb] & $[0,\infty]$ \\ 
 Twist $\tau$      & Number of turns in the magnetic field from one foot point of flux rope to the other [unitless] & $[0,\infty]$\\ \bottomrule 
 \hline
\end{tabular}
\caption{Defining parameters of the FRi3D CME model provided in column 1, with description and units in column 2 and the corresponding range of values in column 3,  as implemented in EUHFORIA}
\label{tab:fri3d_params_description}
\end{table*}

\section{Implementation of FRi3D in EUHFORIA}
\label{sec:Implementation}
\begin{figure}[htb!]
    \centering
    \includegraphics[width=9cm]{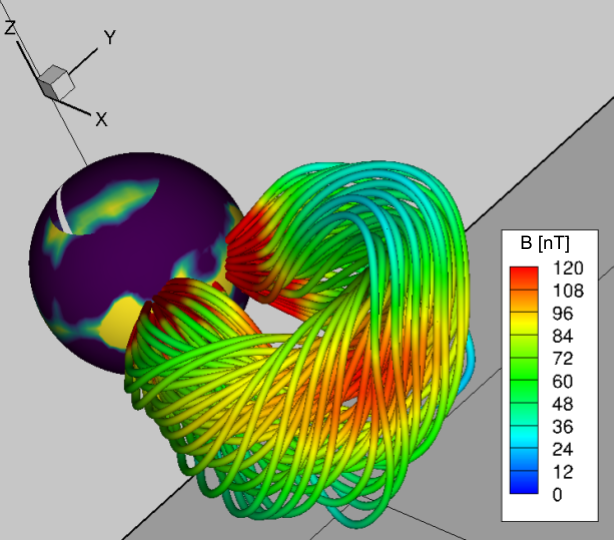}
    
    \caption{A FRi3D flux rope emerging out of EUHFORIA's inner heliospheric boundary at $0.1\;$au. The colour scheme of the field lines is based on the magnetic field strength (B). The field lines are twisted and deformed as per a particular CME geometry and have maximum strength close to the axis which reduces outward (typical Lundquist model characteristics).}
    \label{fig:FRI3D02}
\end{figure}

The cone and spheromak models have disadvantages, to overcome some of them FRi3D was developed. CMEs are injected into EUHFORIA's heliospheric domain after obtaining a steady-state solar wind background in a reference frame co-rotating with the Sun. In this section we will discuss a few important points: a) the incorporation of the FRi3D CME in EUHFORIA; b) the methodology of computing its cross-section (mask) as input; c) the CME leg disconnection technique. A summary of the FRi3D parameters is provided in Table \ref{tab:fri3d_params_description}. Figure~\ref{fig:FRI3D02} depicts a FRi3D flux rope at an early stage of evolution when it is still connected to the EUHFORIA inner boundary. 

\subsection{Geometrical parameters}

A variety of parameters{\color{black}{, as illustrated in Figure~\ref{fig:geo_params}}}, determine the geometry of the FRi3D flux rope such as the angular half-width ($\varphi_{hw}$) and the angular half-height ($\varphi_{hh}$), as compared to the spheromak model whose symmetric radius is defined only by the half-width parameter. The toroidal height ($R_t$), \ie, the heliocentric distance to the apex of the axis, and the poloidal height ($R_p$), \ie, the radius of the cross-section at the apex of the structure, help trace the leading edge of the flux rope shell. The parameters of the flexible outer shell in FRi3D enable the modelling of typical deformations within the inner heliospheric boundary at $0.1\;$au. {\color{black}{The global CME structure including deformations within $0.1\;$au is required to provide an accurate CME cross-section at $0.1\;$au, which in turn affects the geoeffectiveness predictions at $1\;$au \citep{Scolini2018}}}. Pancaking is implemented as the effect of a latitudinal stretch and characterized by an angle $\varphi_p$ which acts as a lateral half-width. The front flattening of the CME experienced immediately after its launch (fast CMEs) can be set with a single coefficient $n$ that adjusts the shape of the axis of the shell. The skew is set by the $\varphi_s$ parameter, which is a rotational transformation about the z-axis. {\color{black}{The deformation parameters in the FRi3D model are illustrated in Figure~\ref{fig:shell_deformation}}}. {\color{black}{\citet{Isavnin2014} showed that a significant amount of the deflection and rotation happens between 30~R$_\odot$ and 1~au. Skew is applicable while fitting flux ropes at large heliocentric distances when the accumulated rotation of the Sun already has a large effect on the global flux rope geometry. For small heliocentric distances, like during the CME injection into EUHFORIA at the inner boundary of $0.1$~au, the skew is negligible and hence, set to zero in this work.}} Apart from the geometry, the central latitude ($\theta$) and longitude ($\phi$) determine the flux rope's position.

The leading edge of the FRi3D CME is defined as the sum of the toroidal height ($R_t$) and the poloidal height ($R_p$). Therefore, the total speed of the CME is defined as a sum of the contribution of height growth due to linear propagation and the increase in the radial cross-section due to pancaking effects. The total 3D speed is thus defined as:

\begin{align}
    v_{3D} &= \frac{d}{dt}(R_t + R_p) = v_{R_t} + v_{R_p}, \label{eqn:vfri}
\end{align}
where 
\begin{align*}
    v_{R_t} = \frac{d}{dt}(R_t) \hspace{0.2cm}  \text{and}   \hspace{0.2cm} v_{R_p} = \frac{d}{dt}(R_p).
\end{align*}

At the inner boundary of the heliospheric module of EUHFORIA, the CME is radially advanced by updating the toroidal height as:
\begin{align}
    R_{t}(t) = R_{t,0} + v_{R_t} \cdot t,
\end{align}
where $R_{t,0}$ is the initial toroidal height during the CME injection at $0.1$~au and $t$ is the time.
Since the poloidal height is implicitly defined as, $R_p = R_t \tan(\varphi_{hh})$, the increment due to the expansion of the cross-section is included self-consistently. $R_p$ is therefore redundant and {\color{black}{uniquely}} defined when $R_t$ and $\varphi_{hh}$ are determined and hence, is not appearing in Table~\ref{tab:fri3d_params_description}, \ref{tab:syn_euh_params} and \ref{tab:euh_params}.

\subsection{Magnetic field parameters}

FRi3D has five magnetic field parameters, viz.\ the tilt, the magnetic flux, the twist, the chirality, and the polarity. {\color{black}{First, a cylinder is populated with parallel field lines.}} The magnetic field configuration of the CME follows the Lundquist model \citep{Lundquist1950} in cylindrical geometry with:
\begin{equation}
B_{\rho} = 0 \text{,} \qquad
B_{\varphi} = B_0 J_1(\alpha \rho) \qquad \text{and} \qquad
B_{z} = B_0 J_0(\alpha \rho) \qquad
\end{equation}



where $\rho$ is the poloidal distance from the axis in cylindrical coordinates, $B_0$ is the strength of the core field, $J_0$ and $J_1$ are the Bessel functions of first and second order{\color{black}{, and $\alpha\rho$ gives the first zero of $J_0$ at the edge of the flux rope ($\alpha$ is a free parameter)}}. 
{\color{black}{The field lines are directed from one footpoint to another by the polarity parameter}}. In the classical Lundquist representation of a flux rope, the twist of the magnetic field lines increases towards the edge of the flux rope, diverging to infinity. As per observations of \citet{Hu2015}, in FRi3D, the magnetic field is initiated with a constant twist ($\tau$), assuming that the foot point of individual field lines does not change. Furthermore, flux conservation is implemented. Although a magnetic field of cylindrical geometry is implemented, the field lines are tapered and bent according to the shell shape. Pancaking and skew transformations are applied to the resultant field line structure.
The aforementioned geometrical and magnetic field parameters can be constrained using remote and in-situ observations (see e.g., Section~\ref{sec:Case_real}). It is also possible to generate synthetic FRi3D CMEs by initializing the parameters based on the range mentioned in Table~\ref{tab:fri3d_params_description} and studying their propagation with EUHFORIA. 

\subsection{Plasma Parameters}
The FRi3D CME is filled with uniform density plasma. In this study, we adopt an observation-based density value coming from the study by \citet{Temmer2021} that suggests an average density of $10^{-17}$~kg~m$^{-3}$ at $0.1\;$au. As it is difficult to constrain the temperature from observations, a uniform CME temperature of $0.8$~MK is considered at the inner boundary \citep{Pomoell2018}. 


\subsection{Mask computation and leg disconnection}
{\color{black}{A mask describes the area where the flux rope structure intersects the inner boundary of the EUHFORIA domain. 
This cross-section is time-dependent due to the gradual propagation of a flux rope through it. When disconnecting the legs from the inner boundary, we gradually reassign the plasma variables of the background solar wind within the cross-section.}}

The distance from the Sun to the CME axis is calculated as \citep[see Eq.~(14)]{Isavnin2016}:
\begin{equation}
    r(\varphi) = R_t\cos^n(a\varphi),
    \label{fri3d_axis_formula}
\end{equation}
where $a = (\pi/2)/\varphi_{hw}$. 
The mask is calculated for $\varphi \ \in \ [-\varphi_{hw}, \varphi_{hw}]$. 
Using Equation~\eqref{fri3d_axis_formula}, the distance from the Sun to a point on the FRi3D flux rope at a certain azimuthal angle of the boundary is computed. 
If it exceeds $0.1\;$au, then this point on the CME has crossed the inner boundary. 
The 3D magnetic field is evaluated at the points where the CME intersects the inner heliospheric boundary. 
Unlike the spheromak model, which is not rooted at the Sun and is hence completely pushed across the inner heliospheric boundary {\color{black}{during the extent of one simulation, the FRi3D flux rope remains rooted at the Sun unless disconnected when specific conditions in the simulation are met. In this work, the FRi3D CME legs are progressively disconnected from the inner boundary point-by-point as soon as the CME speed becomes smaller than the ambient solar wind speed at grid points where the CME is being inserted. Ambient solar wind conditions are then reassigned, and the CME is disconnected, grid cell by grid cell. We note that disconnecting the CME legs from the inner boundary in EUHFORIA simulations is important to avoid an overestimation of the total CME mass and to prevent problems related to the insertion of successive CMEs in the simulation domain. Alternative disconnection mechanisms will be tested in future works.}}

\section{{\color{black}{Comparison of FRi3D and spheromak models in EUHFORIA}}} 
\label{sec:Case_synthetic}

To verify the implementation of FRi3D in EUHFORIA, we perform FRi3D simulations of synthetic (\ie, hypothetical) CME events in a realistic background solar wind. 
We also run simulations with comparable spheromak CMEs, in order to {\color{black}{demonstrate}} FRi3D's performance. 
In this section, we present one such test-case study. 
We consider a CME launched with an intermediate speed along the Sun-Earth line and with an average magnetic field strength to represent an average CME event \citep{Gopalswamy2014}. 
Although the CMEs are synthetic, we choose a realistic background solar wind to have a better propagation environment for the CME. 
The solar wind background is created using the synoptic standard GONG map of 14 May 2020 at 00:14 UT (solar minimum period). 
After obtaining a relaxed solar wind, we design analogous spheromak and FRi3D CMEs to ensure valid testing. {\color{black}{A detailed discussion about the parameters of the spheromak model can be found in \citet{Verbeke2019}}}.


\subsection{Constructing the synthetic CMEs}
\label{density_discrepancy}
In this section, we compare the setup of CME parameters for both models. 

\subsection*{Comparison of FRi3D and spheromak shapes}
In order to compare the simulations with the spheromak and FRi3D CME models, respectively, it must be ensured that, despite different geometry, the CMEs are comparable. We design the synthetic CMEs in such a way that, irrespective of the geometry of the models, they contain a comparable mass. Since we are launching both CMEs with the same speed, the total momentum is also comparable. 
\begin{figure}[!htb]
    \includegraphics[trim={0.5cm 2cm 1cm 0},clip,width=0.5\textwidth]{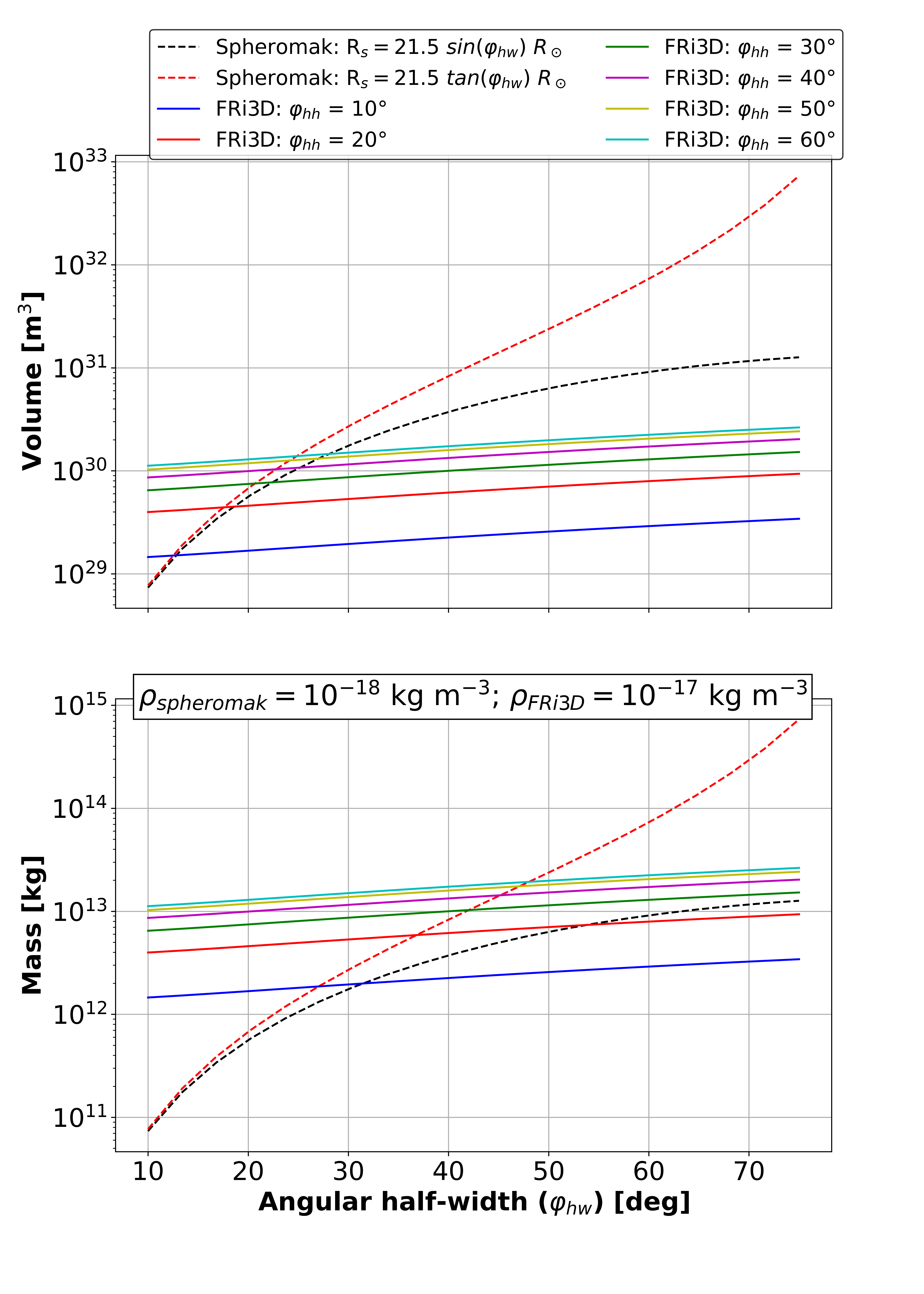}
    \caption{Comparison of volume (top) and mass (bottom) profile of FRi3D vs. half width ($\varphi_{hw}$), for a range of half-heights ($\varphi_{hh}$, values provided above the figure). Two spheromak geometries (sine-geometry: $R_s = 21.5 \sin(\varphi_{hw})$ in red dashed line and tan-geometry: $R_s = 21.5 tan(\varphi_{hw})$ in black dashed line) are over-plotted for comparison. A higher density of $10^{-17}$~kg~m$^{-3}$ based on observations \citep{Temmer2021} is used for the FRi3D CME to make its mass comparable to the spheromak CME, which is simulated with a density of $10^{-18}$~kg~m$^{-3}$ in EUHFORIA. As FRi3D volume depends on both the $\varphi_{hw}$ and $\varphi_{hh}$ values (Appendix A), FRi3D geometries for a range of $\varphi_{hh}$ are shown in solid lines as a function of $\varphi_{hw}$.}
    \label{fig:spheromak_fluxrope_shape}
\end{figure}
\begin{figure*}[!htb]
    \centering
    \includegraphics[trim={0.9cm 2.5cm 0.4cm 0.5cm},clip,width=19cm]{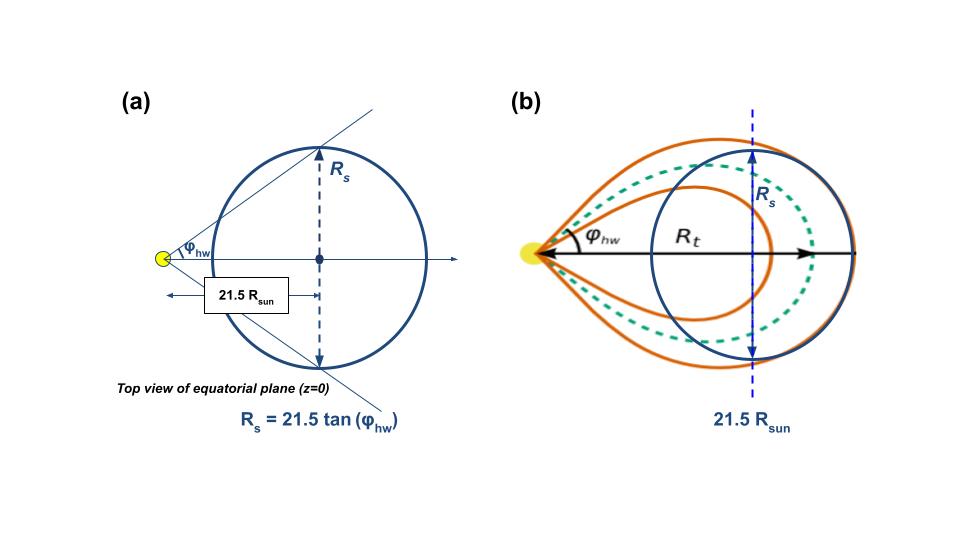}
    \caption{{\color{black}{Schematic c}}omparison of geometry of the spheromak and FRi3D models. (a) The spheromak (tan-geometry) CME when it is halfway through the inner boundary; (b) The comparison of the FRi3D shape with the spheromak at the same instant. The circle (with radius $R_s$) shows the spheromak CME cross-section as seen from the top of the equatorial plane, compared to the FRi3D CME. The blue dashed vertical line shows the {\color{black}{position of the}} inner boundary. The spherical structure of the spheromak flux rope fails to capture the CME legs that are modelled by the extended geometry of the FRi3D flux rope.
}
    \label{fig:FRI3Dshape}
\end{figure*}
The volume of FRi3D (see \ref{sec:FRi3D_volume} for details) is plotted in the top panel of Figure~\ref{fig:spheromak_fluxrope_shape} for a range of $\varphi_{hh}$ and $\varphi_{hw}$. It is compared to the spheromak volume for two geometries, $R_s = 21.5 \tan (\varphi_{hw})$ (tan-geometry, hereafter) and $R_s = 21.5 \sin (\varphi_{hw})$ (sine-geometry, hereafter), where $R_s$ is the spheromak radius.
The spheromak volume is up to three orders of magnitude higher for the wider CMEs using the tan-geometry, while the volume with sine-geometry lies within an order of magnitude of FRi3D volume. Although the sine-geometry might seem like a better choice to make the spheromak model comparable to FRi3D, studies by \citet{Scolini2018,Scolini2019} show that the spheromaks with tan-geometry inserted at the EUHFORIA's inner boundary provide the best predictions of ICME properties and their geoeffectiveness at Earth. Therefore, we use the tan-geometry of the spheromak in this study. However, in order to match the mass in the spheromak CME, the conventional CME density of $10^{-18}$~kg~m$^{-3}$ in EUHFORIA has to be increased for the low volume FRi3D CME. Figure~\ref{fig:FRI3Dshape} compares cross-sections of a FRi3D CME and a spheromak CME with tan-geometry, when the CMEs are initialized at the inner heliospheric boundary. Figure~\ref{fig:FRI3Dshape}(b) clearly illustrates how the spheromak fails to model the CME legs.

 Recent statistical study on deriving CME density from the observations by \citet{Temmer2021} shows that the average CME mass density ($\rho$) lies close to $10^{-17}$~kg~m$^{-3}$ at $0.1\;$au. This conclusion is consistent with our hypothesis of increasing the density of the CMEs modelled with FRi3D. 
 Taking $\rho_{FRi3D}=10^{-17}$~kg~m$^{-3}$ and $\rho_{spheromak}=10^{-18}$~kg~m$^{-3}$, the mass of CMEs with $\varphi_{hw} \in [20,60]$ is comparable for both models (in the range of $10^{12}-10^{14}$~kg), as shown in the bottom panel of Figure~\ref{fig:spheromak_fluxrope_shape}. Based on this analysis, we choose the $\varphi_{hw} = 45\degree$ and $\varphi_{hh} = 20\degree$ for constructing our synthetic CMEs. With this realisation, we perform three EUHFORIA simulations: the spheromak model (tan-geometry), FRi3D ($\rho=10^{-17}$~kg~m$^{-3}$) and FRi3D$\_{\text{low}}$ ($\rho=10^{-18}$~kg~m$^{-3}$). Two FRi3D simulations are performed to illustrate the effect of CME mass on its evolution in EUHFORIA's heliospheric domain. All the simulation parameters are listed in Table~\ref{tab:syn_euh_params}. 


\subsection*{Speed at launch}
The propagation speed of a CME can be decomposed as
\begin{equation}
    v_{3D} = v_{rad} + v_{exp},
\end{equation}
where 
\begin{align*}
    v_{rad} = \frac{1}{1 + \kappa}v_{3D} \hspace{0.25cm} \text{and} \hspace{0.25cm} v_{exp} = \frac{\kappa}{1 + \kappa}v_{3D}
\end{align*}
 are the radial and expansion speed of the CME, which can be observationally constrained as a function of the GCS aspect ratio parameter $\kappa$ \citep{Scolini2019}.  {\color{black}{$\kappa$ is the ratio of the CME size in two orthogonal directions that is set to a constant to enable self-similar expansion of the CME  \citep[][]{Thernisien2006,Thernisien2009}}}. Following the results of \citet{Verbeke2019} and \citet{Scolini2019}, the spheromak should be launched by setting the radial speed as the injection speed, since the expansion speed is self-consistently taken care of by the model. The FRi3D launch speed is set equal to its toroidal speed $v_{R_t}$ (see Eq.~\eqref{eqn:vfri}). However, we recommend setting the launch speed as obtained from the CME reconstruction, done using FRi3D or GCS, pertaining to the chosen model. Although the spheromak's $\kappa$ and FRi3D's $\varphi_{hh}$ seem to be analogous, they can be fitted differently in both the reconstructions. This introduces a variance in $v_{rad}(\kappa)$ and $v_{R_t}(\varphi_{hh})$. {\color{black}{In this section, we launch the spheromak and the FRi3D CME with the same injection speed, i.e., the toroidal speed of FRi3D is assumed to be equivalent to the radial speed of the spheromak. As we do not take into consideration the variation in the injection speeds of both models due to different expansion factors, we do not compare the arrival time of the CMEs at Earth. The emphasis is put more on the comparison of the behaviour of other plasma properties at Earth}}. 
 
 
\subsection*{FRi3D toroidal height}
The FRi3D CME is launched from the inner boundary with the configuration such that the leading edge of the self-similarly expanded CME touches the inner boundary. The toroidal height during injection of the CME at the inner boundary is calculated by inverting the following relation:
\begin{equation}
    R_t + R_p = 21.5 R_\odot.
    \label{eqn:Rt}
\end{equation}

\subsection*{Magnetic field parameters}
The implementation of the spheromak in EUHFORIA requires information on the toroidal magnetic flux, whereas FRi3D uses total magnetic flux as an input parameter. 
For the synthetic case under consideration, we try to match the order of the magnitude of flux in both models. 
Both models are given the same chirality (left-handed). In the case of the spheromak, the orientation of the CME is completely determined using the chirality and the flux rope tilt angle. Along with chirality, the polarity parameter in FRi3D determines its full-fledged orientation. 
We consider a left-handed chirality and a west-to-east polarity for the flux rope. The tilt of the spheromak CME is chosen by aligning the axis of symmetry of the spheromak with the magnetic axis of the FRi3D model. 
The comparison of parameters, for two considered models, is presented in Table \ref{tab:syn_euh_params}.

\begin{table}
\begin{tabular}{ |p{3cm}||p{2.5cm}|p{2.2cm}|p{2.2cm}|  }
 \hline
 \multicolumn{3}{|c|}{\textbf{Input parameters}} \\
 \hline
 CME model   &  Spheromak & FRi3D (FRi3D$\_{\text{low}}$) \\
 \hline
  \multicolumn{3}{|c|}{{Geometrical}} \\
 \hline
 Insertion time   & {\color{black}{2020-05-14T13:52:00}} & {\color{black}{2020-05-14T13:52:00}} \\
 CME Speed   & $650$~km~s$^{-1}$ &  $650$~km~s$^{-1}$\\
 Latitude    & $0\degree$ & $0\degree$ \\
 Longitude   & $0\degree$ & $0\degree$ \\
 Half-width  & - & $45\degree$ \\
 Half-height & - & $20\degree$ \\
 Radius      & $21.5$ R$_\odot$ & - \\
 Toroidal height & - & $15.76$ R$_{\odot}$\\
\hline
 \multicolumn{3}{|c|}{{Magnetic field}} \\
 \hline
 Chirality    & $-1$ & $+1^*$ \\ 

 Polarity    & - & $+1$ \\
 Tilt        & $-90\degree$ & $0\degree$ \\
 Toroidal magnetic flux & $1 \cdot 10^{13}$~Wb & - \\
 Total magnetic flux & - & $1\cdot10^{13}$~Wb \\
 \hline
 \multicolumn{3}{|c|}{{Deformation}} \\
 \hline
 Flattening  & - & 0.5\\
 Pancaking   & - & 0.5\\
 Twist       & - & 2.0\\
\hline
 \multicolumn{3}{|c|}{{Plasma parameters}} \\
\hline
 Mass density     & $10^{-18}$~kg~m$^{-3}$ & $10^{-17}$~kg~m$^{-3}$ ($10^{-18}$~kg~m$^{-3}$) \\
 Temperature & $0.8 \cdot 10^6$~K & $0.8 \cdot 10^6$~K \\
 \hline
 {\color{black}{Arrival time at Earth}} & {\color{black}{2020-05-16T19:03}} & {\color{black}{2020-05-16T10:33 (2020-05-16T13:13)}} \\    
 \hline
\end{tabular}
\caption{CME parameters used in the EUHFORIA simulations of the synthetic event employing the spheromak and the FRi3D model. Two FRi3D simulations are performed that are identical in all parameters except mass density - FRi3D (high mass density) and FRi3D$\_{\text{low}}$ (low mass density). \\
$^*$FRi3D chirality is implemented with an opposite convention i.e., -1 for right-handedness and +1 for left-handedness.}
\label{tab:syn_euh_params}
\end{table}

\subsection{EUHFORIA numerical setup}
In this section, the simulation setup of EUHFORIA (version 2.0) is discussed.
The semi-empirical coronal model of EUHFORIA is set to the default configuration prescribed by \citet{Pomoell2018}. 
The computational domain of the heliospheric part extends from $0.1\;$au to $2.0\;$au in the radial direction, $\pm 60 \degree$ in latitude, and covers the full $360\degree$ in the longitudinal direction. We use $256$ cells in the radial direction and the latitudinal and longitudinal resolution is set to  $4\degree$ and $2\degree$, respectively. 
Additional virtual spacecraft are placed at different radial distances (from $0.1\;$au to $2.0\;$au at the interval of $0.1\;$au) and longitudes ($-65\degree$ to $65\degree$ at an interval of $5\degree$) to compare the spheromak and the FRi3D models at CME flanks. 

\begin{figure*}[htb!]
    \centering
    \subfloat[FRi3D: Radial velocity]{{\includegraphics[width=0.49\textwidth]{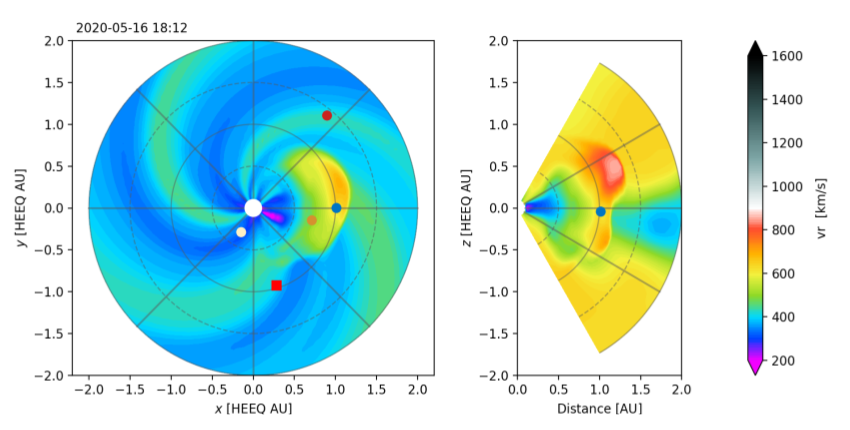} }} 
    \subfloat[Spheromak: Radial velocity]{{\includegraphics[width=0.49\textwidth]{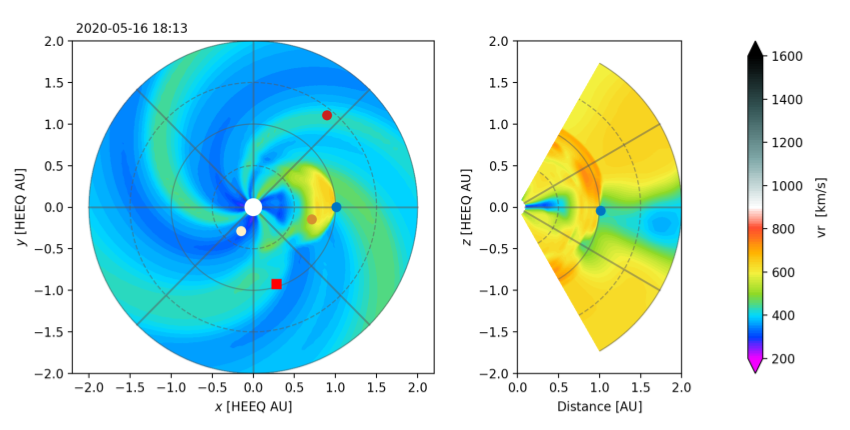} }}\\
    \subfloat[FRi3D: Scaled number density]{{\includegraphics[width=0.49\textwidth]{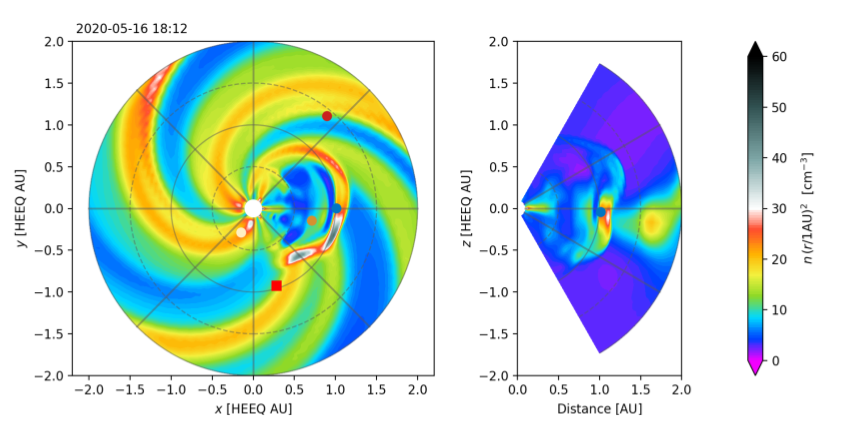} }} 
    \subfloat[Spheromak: Scaled number density]{{\includegraphics[width=0.49\textwidth]{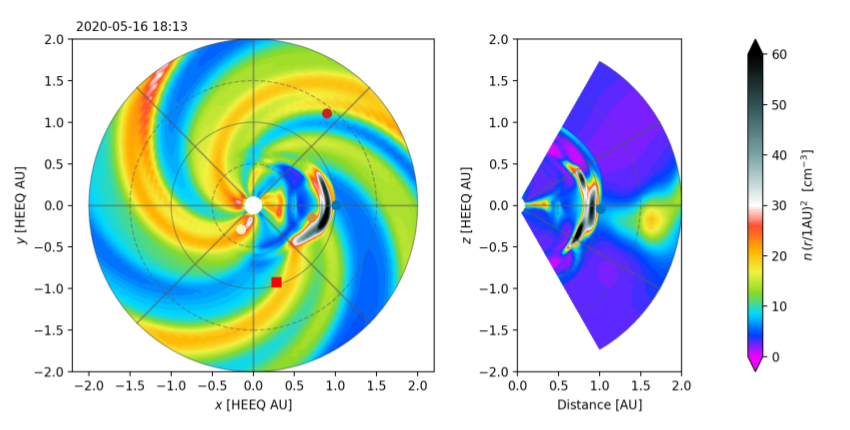} }}\\
    \subfloat[FRi3D: Co-latitudinal magnetic field component]{{\includegraphics[width=0.49\textwidth]{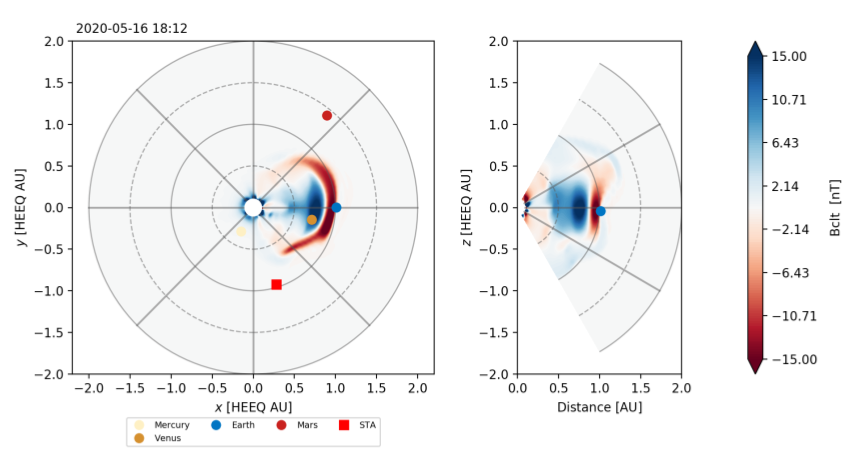} }} 
    \subfloat[Spheromak: Co-latitudinal magnetic field component]{{\includegraphics[width=0.49\textwidth]{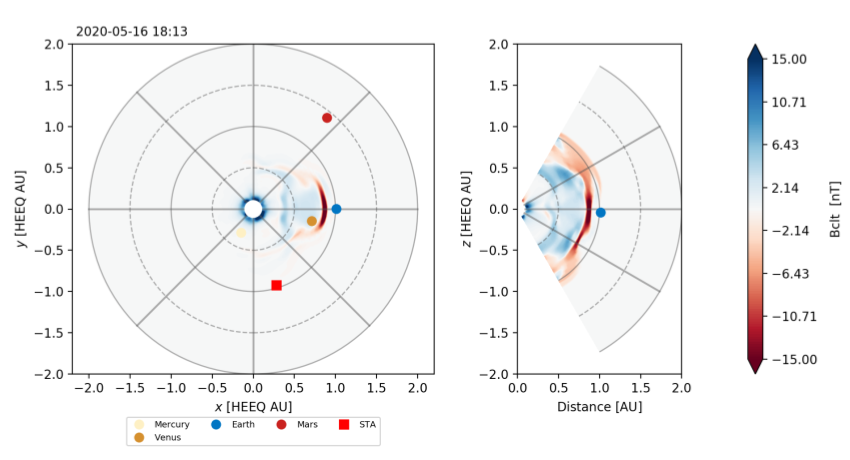} }}
    \caption{EUHFORIA simulation snapshots on May 16, 2020 at 18:12 upon arrival {\color{black}{at Earth}} of the synthetic CME, injected at the inner boundary on May 14, 2020 at 13:52 UT. \textit{Top to bottom}: radial velocity ($v_r [\text{km s}^{-1}]$), scaled number density ($n\cdot(r/1~au)^2 [\text{cm}^{-3}]$) and co-latitudinal magnetic field component ($B_{clt} = -B_z [\text{nT}]$) in the equatorial {\color{black}{(panels with $x$ [HEEQ AU] - $y$ [HEEQ AU] axes)}} and the meridional {\color{black}{(panels with $x$ [HEEQ AU] - $z$ [HEEQ AU) axes)]}} planes of EUHFORIA's heliospheric domain are plotted for the FRi3D simulation with $\rho=10^{-17}$~kg~m$^{-3}$ (left panel, (a), (c) and (e)) and the spheromak simulation (right panel, (b), (d) and (f)). The extended longitudinal part of the FRi3D CME can be distinctly observed in the scaled density {\color{black}{(panel c, left plot)}} and $B_z$ equatorial plots {\color{black}{(panel e, left plot)}} due to the presence of CME legs. The spheromak CME, although has a latitudinal coverage as seen in the meridional plane, lacks a longitudinal extension. {\color{black}{This illustrates the potential impact of FRi3D at the virtual spacecraft placed around the Sun-Earth line.}}}
    \label{fig:20200514_eq_mer_plot}
\end{figure*}

A snapshot of the EUHFORIA heliospheric domain is plotted in Figure~\ref{fig:20200514_eq_mer_plot}. A comparison of radial velocity, scaled density, and B$_z$ between the FRi3D (left) and the spheromak (right) simulations is illustrated. The larger longitudinal extent of the FRi3D flux rope is evident from the equatorial profile of {\color{black}{$B_z$ (Co-latitudinal component) and scaled number density}}. Further quantitative comparative analysis is done in the next section.


\begin{figure*}[htb!]
    \centering
    \includegraphics[trim={0.2cm 0.3cm 0 0.3cm},clip,width=17cm]{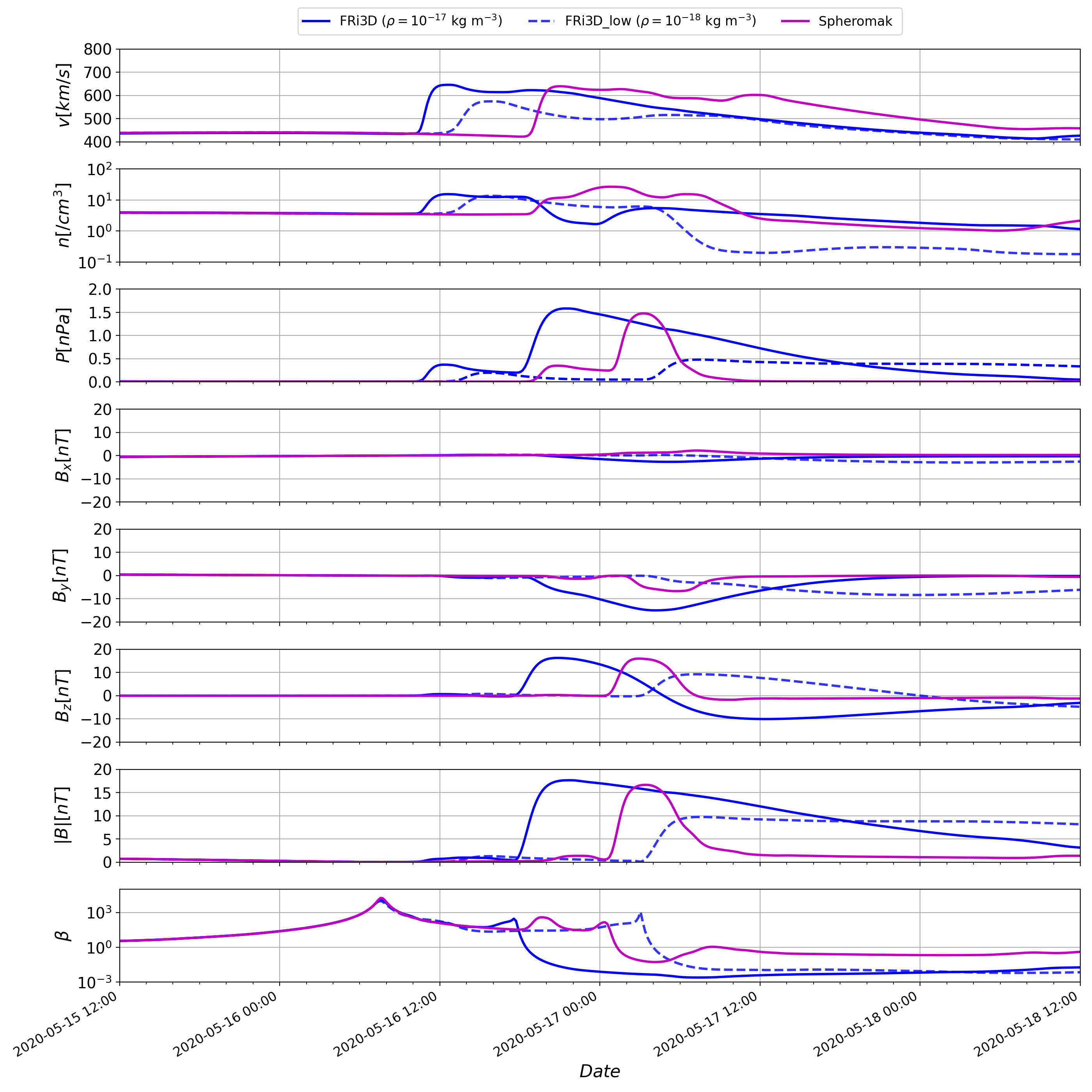} 
    \caption{Plasma and magnetic field properties {\color{black}{at Earth as a function of time}}, of the three synthetic CMEs, that were inserted at the EUHFORIA inner boundary on 14 May 2020 with initial parameters as discussed in Table~\ref{tab:syn_euh_params}. \textit{from top to bottom}: speed ($v$), number density ($n$, in log scale), total pressure (sum of thermal and magnetic pressure, $P$) and plasma beta ($\beta$, in log scale); $B_x$, $B_y$, $B_z$ components in GSE coordinates and magnetic field strength ($B$) are plotted for FRi3D, FRi3D$\_{\text{low}}$ and the spheromak simulations. It can be observed that $n$ in the FRi3D$\_\text{low}$ simulation drops below $1/\text{cm}^3$ and the magnetic field components are even lower than in the case of the spheromak. The magnetic field components are significantly improved and {\color{black}{the rotation of the magnetic field in $z$ direction}} is distinct in the FRi3D simulation with higher density.}
    \label{fig:syn_timeseries}
\end{figure*}

\begin{figure}[htb!]
    \centering
    \includegraphics[trim={0.6cm 0.75cm 0.8cm 0},clip, width=0.5\textwidth]{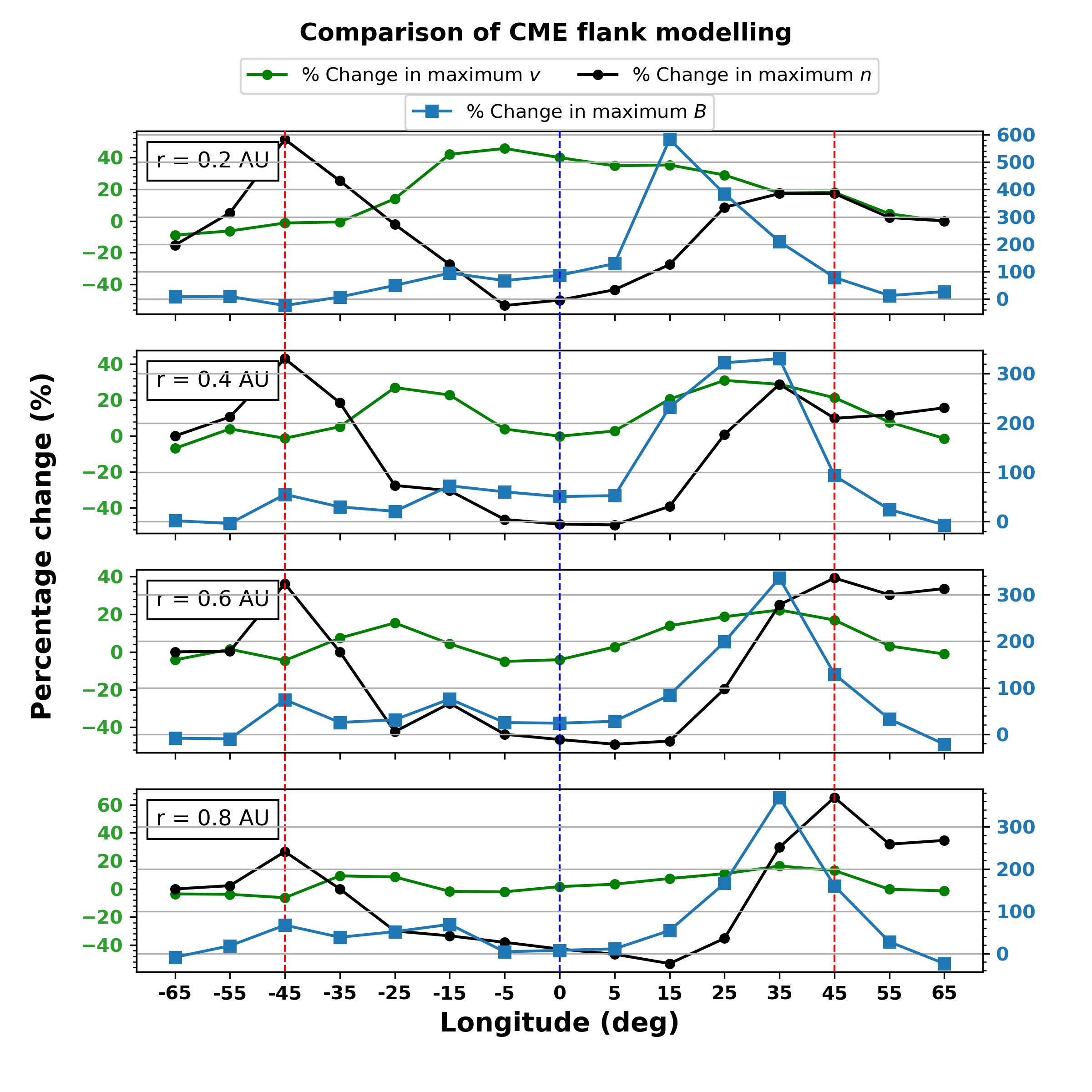}
    \caption{Comparison of the FRi3D and the spheromak maximal speeds and magnetic field strengths at different heliocentric distances and multiple longitudes at a particular radial distance. {\color{black}{Left side of the y-axis denotes the \% change in speed and number density, and the right side represents the \% change in magnetic field strength respectively. The vertical red dashed lines mark the position of the initial angular half-width at $\pm45\degree$ longitude and the vertical blue dashed line indicates the longitude of the launch of the CME \ie, the Sun-Earth line.}}} 
    \label{fig:FRI3D05}
\end{figure}

\subsection{Results}
The FRi3D and the spheromak CMEs evolve differently due to their entirely different magnetic field topologies, which is reflected in the solar wind plasma properties at Earth. Moreover, discrepancies in the CME arrival times and other plasma parameters depend on the different approaches used to initialize the FRi3D flux rope geometry compared to the spherical spheromak geometry, as detailed in Section~\ref{density_discrepancy}. The evolution of the plasma and magnetic field properties at Earth are shown in Figure~\ref{fig:syn_timeseries}. 
The solid and dashed blue profiles represent the FRi3D CMEs with high and low density, respectively. The spheromak properties are plotted in magenta. The trends in the plasma profiles are similar and have similar magnitudes (for a head-on CME impact) at Earth. 
The {\color{black}{CMEs in the}} FRi3D and the FRi3D$\_{\text{low}}$ simulations arrive earlier than {\color{black}{the CME in}} the spheromak simulation by about $8$ and $10$ hours respectively. As discussed previously, it may be necessary to launch the CMEs with speed derived from observations while modelling real events. The magnetic field components at $1\;$au obtained with FRi3D have significantly larger values and the rotation of {\color{black}{the magnetic field vector in the $z$-direction}} is distinctly captured as compared to the spheromak. Furthermore, Figure~\ref{fig:syn_timeseries} illustrates that the plasma inside the flux rope of the low-density FRi3D CME is extremely rarefied, with number densities dropping below $1/\text{cm}^3$. 
This is because the CME undergoes over-expansion after its insertion at the inner boundary, due to its strong internal magnetic field and the low initial density inside the CME. Due to this expansion, both the density and the magnetic field are significantly decreased by the time the CME arrives at Earth. By increasing the initial density of the CME, the inertia of the plasma inside the CME is increased, which counteracts the magnetic expansion, and hence restricts the excessive reduction of the magnetic field strength and density. 
As a result, the density profile along with other plasma properties, are significantly different when launching FRi3D with a higher initial density. The magnetic field strength is almost twice as high for the FRi3D as compared to the FRi3D$\_{\text{low}}$ simulation.

{\color{black}{In summary, injecting the spheromak and FRi3D CMEs with comparable mass resulted in comparable magnitudes of physical properties between the two models. Secondly, initializing a FRi3D CME with the density constrained by observations solved the problem of low number density inside the CME as it propagated away from the Sun. Hence, we propose that in our simulations, the total mass contained in a CME is a more important parameter than its local density. Therefore, we conclude that using a standard mass density {\color{black}{(reported in the statistical study by Temmer et al., 2021) in the order}} of $10^{-17}$~kg~m$^{-3}$ for initializing the FRi3D CMEs in EUHFORIA is optimal.}}

One of the main reasons for coupling the FRi3D model with EUHFORIA is {\color{black}{the presence of CME legs in}} its global geometry. Hence, we perform a preliminary analysis to test FRi3D's performance in modelling plasma properties at the CME flanks. In future studies, we will validate th{\color{black}{e same}} by modelling observed cases of flank encounters. In this analysis, a number of virtual spacecraft are placed at different heliocentric distances between the inner boundary and Earth, separated radially by $0.1$~au. Similarly, virtual spacecraft are also placed at different longitudes at an interval of $5\degree$ from $-65\degree$ to $65\degree$ around the Sun-Earth line {\color{black}{at the above-mentioned}} radial distances. As the CMEs are launched along the Sun-Earth line, this arrangement of the spacecraft facilitates the study of the impact of the CME flanks. To quantify the differences in the maximal speed ($v$), number density ($n$), and magnetic field strength ($B$) between the FRi3D and the spheromak CMEs, we  define $\Delta v_{\max}$, $\Delta n_{\max}$ and $\Delta B_{\max}$ as
\begin{equation}
    \Delta v_{\max} = \frac{\max(v_{\text{FRi3D}}) - \max(v_{\text{spheromak}})}{\max(v_{\text{spheromak}})} \times 100
\end{equation}

\begin{equation}
    \Delta n_{\max} = \frac{\max(n_{\text{FRi3D}}) - \max(n_{\text{spheromak}})}{\max(n_{\text{spheromak}})} \times 100
\end{equation}

\begin{equation}
    \Delta B_{\max} = \frac{\max(B_{\text{FRi3D}}) - \max(B_{\text{spheromak}})}{\max(B_{\text{spheromak}})} \times 100
\end{equation}

Figure~\ref{fig:FRI3D05} shows $\Delta v_{max}$, {\color{black}{$\Delta n_{max}$}} and $\Delta B_{max}$ obtained with virtual spacecraft at different radial and longitudinal positions. The vertical red dashed lines mark the position of the initial angular half-width at $\pm45\degree$ longitude and the vertical blue dashed line indicates the longitude of the launch of the CME \ie, the Sun-Earth line. {\color{black}{The key observations of this analysis are three-fold. (1)}} For head-on impact, within $\pm10\degree$ shift from Earth, the enhancement in speed is within $5 \%$. However, when the observer is located at the flanks of the CME, upon impact, the FRi3D model yields an enhancement {\color{black}{in speed up to $50\%$ close to the Sun and up to $20\%$ at $0.8$~au. {Note that,} the maximum speed that we derive from the time series, belongs to the sheath region ahead of the CME. {\color{black}{The characteristics of the sheath region depend}} on the propagation of the CME relative to the solar wind and its expansion (model-dependent) \citep{Siscoe2008,Kaymaz2006}. {\color{black}{As we compare the two CME models, it must be highlighted that}} along with the contribution of FRi3D's extended geometry to the speed enhancement, upon using FRi3D over the spheromak model, there can be a contribution from the way the flux rope model drives the sheath ahead of it \citep{Kilpua2017,Russell2002}. {\color{black}{(2)}} The trend in density enhancement is opposite, i.e., the spheromak density is higher than FRi3D density close to the CME nose and within the half-width angle. In FRi3D, the density is spread throughout the extended flux rope whereas, in the spheromak, the density is concentrated close to the nose of the CME {\color{black}{(see, Figure~\ref{fig:20200514_eq_mer_plot}(d))}}. With increasing heliocentric distance, {\color{black}{the density modelled by FRi3D exceeds that of the spheromak}}, beyond 35$\degree$ and 30$\degree$ in the western and eastern flanks respectively as shown in Figure~\ref{fig:FRI3D05}. {\color{black}{(3)}} Although the maximum $v$ corresponds to the sheath region, the maximum $B$ belongs to the magnetic cloud. The enhancement is significant in the magnetic field strength at the flanks. It goes up to $100\%$ on the eastern side of the Sun-Earth line (negative longitudes). On the western side (positive longitudes), the enhancement goes up to $600\%$ at $0.2$~au and reduces to about $400\%$ at $0.8$~au. The asymmetric CME shape and the noticeable difference in the plasma characteristics of a CME, especially in the case of the magnetic field, might be due to the interaction with a high-speed stream west of the CME (see Figure~\ref{fig:20200514_eq_mer_plot}(a))}. When the eastern flank of the CME enters the high-speed stream, i.e., a less dense region (see Figure~\ref{fig:20200514_eq_mer_plot}(c)), it expands asymmetrically into the stream and this results in the enhancement of plasma properties at the position of a virtual spacecraft on the eastern flank.} 
$\Delta B_{max}$ starts to drop close to the flank $\sim 45\degree$ and goes to zero near $\pm60\degree$. {\color{black}{This indicates the start of the region where there is no influence from the CMEs and, for both models, the magnetic field strength (given by the background solar wind model) is therefore identical.}} 



\section{Case study: 12 July 2012 event}
\label{sec:Case_real}
After the demonstrating the implementation of FRi3D in EUHFORIA using a synthetic event, in this section we study an observed CME {\color{black}{and constrain its parameters at $0.1\;$au using remote-sensing and in-situ observations.}} 
We analyzed the Earth-directed halo  CME that erupted from the NOAA AR 11520, on 12 July 2012. This fast CME caused significant geomagnetic disturbances (minimum $Dst$ of $-139$~nT) at Earth. {\color{black}{This is a textbook event that was studied by many authors, see e.g. \citet{Hu2016}\citet{Gopalswamy2018}. We have selected this event for the following reasons: (1) its clear flux rope signature recorded in-situ; (2) availability of multi-viewpoint {\color{black}{coronagraph (remote-sensing)}} observations for reconstruction; (3) a clear {\color{black}{association}} between the eruption source region (on 12 July 2012), CME in the coronagraphic field of view and the ICME that arrived at Earth on 14 July 2012}}. A detailed description of the event and the EUHFORIA simulation can be found in \citet{Scolini2019}. The strong geomagnetic impact of this event, due to its long-lasting negative $B_z$ component of the interplanetary magnetic field (IMF),  was unexpected and poorly forecasted \citep[][]{Webb2017}.

In \citet{Scolini2019}, the geometrical parameters for the spheromak CME are derived using the GCS model on the white light images of STEREO spacecraft. The magnetic field parameters are derived from the remote observations of the active region. The helicity of the flux rope is derived from the pre-eruption AIA $94 \angstrom$ observations of the EUV sigmoids. As the spheromak implementation in EUHFORIA requires only the chirality (i.e., the sign of the helicity), it is determined by whether it is a forward or reverse sigmoid \citep[]{Palmerio2017}. Flux rope tilt is determined from the orientation of the source region polarity inversion line \citep[PIL,][]{Marubashi2015} with the assumption of no further rotation in the lower corona. Finally, the magnetic field flux is computed employing a modified version of the FRED method \citep[][]{Gopalswamy2017}. We reproduce the spheromak simulation using the parameters prescribed by \citet{Scolini2019} to set a reference for the performance of FRi3D and {\color{black}{demonstrate}} the effect of employing the novel CME model in EUHFORIA. In addition, these parameters act as a reference to check the validity of the parameters we derive using the FRi3D tools.

\subsection{FRi3D CME input parameters}
The FRi3D CME parameters are derived independently from remote and in-situ data observations for a clear and unambiguous fitting. 
\subsection*{FRi3D forward modelling tool}
\begin{figure}[!htb]
    \centering
    \includegraphics[width=9cm]{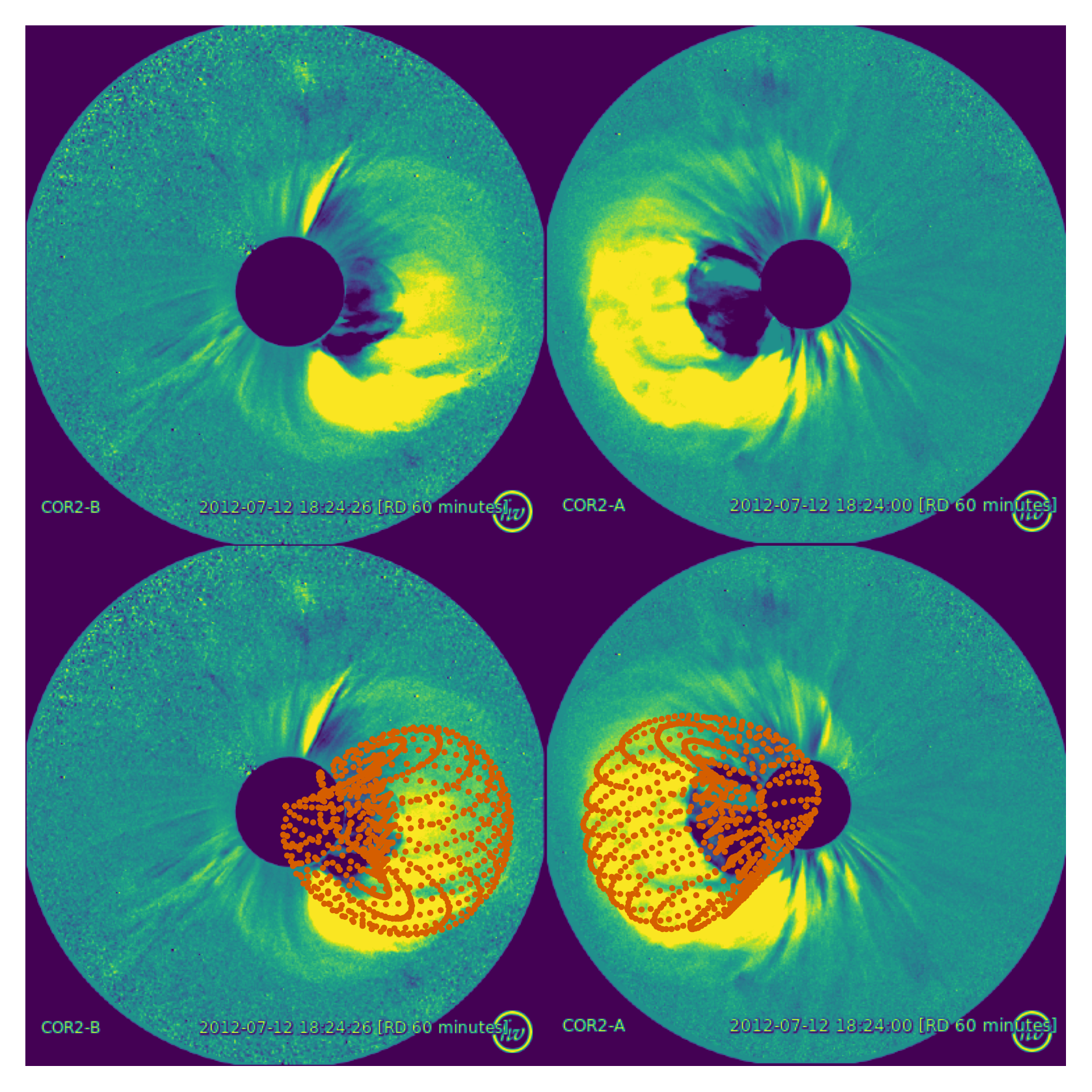} 
    \caption{Multi-viewpoint coronagraph fitting of the CME that erupted on 12 July 2012, by employing FRi3D flux rope to the white-light observations of COR2A (STEREO-A) and COR2B (STEREO-B). {\color{black}{Top figures are shown as a reference without the flux rope fitting.}}}
    \label{fig:FRI3Dcor}
\end{figure}

The forward modelling tool of FRi3D \citep{Isavnin2016} is applied to the remote multi-viewpoint observations from STEREO-A and STEREO-B for the 3D reconstruction. {\color{black}{The separation of STEREO-A and STEREO-B with Earth was $120\degree$ {\color{black}{(on the west)}} and $115\degree$ {\color{black}{(on the east)}}, respectively, on the day of eruption. The CME was observed by LASCO C2 only during its early evolution phase (data available only at 17:12 UT), while LASCO C3 does not have observations of this event. {\color{black}{Therefore, w}}e used just the viewpoint observations from the two STEREOs to track the evolving CME up to $\sim 0.05$~au}}. 
One snapshot of the fitting is demonstrated in Figure~\ref{fig:FRI3Dcor} and the geometrical parameters are listed in Table~\ref{tab:euh_params}. Latitude, longitude, and half-width derived with FRi3D fitting agree closely with the GCS results obtained by \citet{Scolini2019}. Additionally, the half-height, flattening, and pancaking parameters are determined using this technique within the inner heliospheric boundary during the early evolution of the CME. 
Toroidal height is calculated by inverting the Eqn.~\ref{eqn:Rt}. The injection speed of FRi3D corresponds to its toroidal speed which is computed by fitting the FRi3D flux rope to sequential coronagraphic images and computing the rate of change of $R_t$. 
As $R_p$ explicitly depends on $R_t$, the expansion of the cross-section is self-consistently manifested. 
The CME expansion speed is controlled by the aspect ratio in GCS and the half-height in FRi3D. The contribution of the expansion speed can be different in these two models, due to different geometrical constructions. 
For this event, the FRi3D toroidal speed (injection speed) is $\sim$~100~km s$^{-1}$ smaller than the spheromak radial speed obtained from GCS fitting. {\color{black}{Hence, the same CME is injected with different speeds in the two EUHFORIA simulations.}}

\begin{figure}[!htb]
    \centering
    \includegraphics[trim={0.6cm 0.3cm 0.2cm 0.8cm},clip, width=0.5\textwidth]{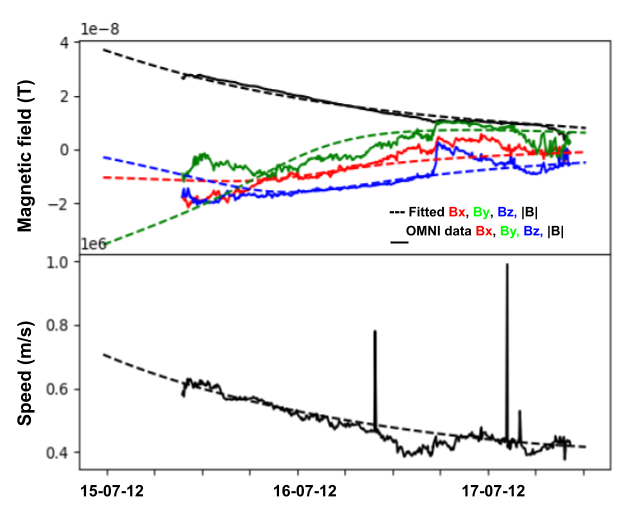}
    \caption{Best numerical fitting of FRi3D model to the in-situ measurements by WIND spacecraft in HEEQ coordinates, of the ICME, launched from the Sun on 12 July 2012 compared to the observed data. The FRi3D model estimated magnetic field components and the total magnetic field (\textit{top}), and the speed (\textit{bottom}) (in dashed line) are compared to the measured data (in solid line).}
    \label{fig:FRI3Dinsitu}
\end{figure}

\subsection*{FRi3D in-situ tool}
The ICME parameters are also obtained by a numerical fitting of the FRi3D CME to the in-situ measurements from the WIND spacecraft at $1\;$au using a differential evolution algorithm \citep{Storn1997}. {\color{black}{Further details of this tool can be found in \citet{Isavnin2016}}}. 
This fitting procedure is performed multiple times reducing the offset between real and modelled data and ensuring uniqueness of convergence. 
As we assume that the magnetic field properties like the flux and twist remain mostly unchanged, we constrain those parameters from the in-situ fitting obtained at $1\;$au. 
While parameters like chirality and polarity of the flux rope from in-situ agree with the remote observations \citep[see e.g.,][]{Scolini2019}, geometrical parameters derived from in-situ fitting differ from the values constrained remotely. 
This can be attributed to the fact that the fitting is done at $1.0\;$au and the flux rope has undergone changes like deflection and rotation upon interaction with heliospheric transients during its evolution, hence resulting in different geometrical parameters as compared to $0.1\;$au \citep[][]{Manchester2017,Kay2013,Lugaz2011d,Kilpua2009,Isavnin2014}. Therefore, we use the remote observations to constrain the geometrical parameters, and in-situ observations to constrain the magnetic field parameters at the inner boundary. 

\begin{table}
\begin{tabular}{ |p{3cm}||p{2.5cm}|p{2.2cm}|p{2.2cm}|  }
 \hline
 \multicolumn{3}{|c|}{\textbf{Input parameters}} \\
 \hline
 CME model   &  Spheromak & FRi3D \\
 \hline
 Insertion time   & 2012-07-12T19:24 & 2012-07-12T19:02 \\
 CME Speed   & $763$~km~s$^{-1}$ &  $664$~km~s$^{-1}$\\
 Latitude    & $-8\degree$ & $-8\degree$ \\
 Longitude   & $-4\degree$ & $-4\degree$ \\
 Half-width  & - & $38\degree$ \\
 Half-height & - & $36.8\degree$ \\
 Radius      & $16.8$~R$_\odot$ & - \\
 Toroidal height & - & $12.29$~R$_\odot$ \\
 Flattening  & - & $0.3$\\
 Pancaking   & - & $0.44$\\
 Skew        & - & $0$\\
 \hline
 Chirality    & $+1$ & $-1^*$ \\ 
 Polarity    & - & $-1$ \\
 Tilt        & $-135 \degree$ & $45 \degree$ \\
 Toroidal magnetic flux & $1 \cdot 10^{14}$~Wb & - \\
 Total magnetic flux & $2.4 \cdot 10^{14}$~Wb & $0.5 \cdot 10^{14}$~Wb \\
 Twist       & - & $1.0$\\ 
 \hline
 Mass density     & $1 \cdot 10^{-18}$~kg~m$^{-3}$ & $1 \cdot 10^{-17}$~kg~m$^{-3}$ \\
 Temperature & $0.8 \cdot 10^6$~K & $0.8 \cdot 10^6$~K \\

 \hline
 Arrival time at Earth & 2012-07-14T19:23 & 2012-07-14T20:53\\
 \hline
\end{tabular}
\caption{CME input parameters used in EUHFORIA simulations of 20120712 event employing the spheromak and the FRi3D model.\\
$^*$FRi3D chirality is implemented with an opposite convention i.e., -1 for right-handedness and +1 for left-handedness.}
\label{tab:euh_params}
\end{table}


As chirality, polarity, and tilt could be determined for this event from clear remote-sensing observations using the methodologies mentioned above, we fix those parameters and reduce the number of free parameters for the numerical fitting algorithm. 
The in-situ data fitting results at $1\;$au are shown in Figure~\ref{fig:FRI3Dinsitu}. The algorithm computes the offset between the observed and the fitted magnetic field ($\delta B$) and speed ($\delta v$) at every iteration. The goodness of fitting is inversely proportional to the sum of $\delta B$ and $\delta v$. 
For the convergence of the algorithm with the highest fitness, the total magnetic flux obtained is $0.5\cdot10^{14}$~Wb ($\phi_{FRi3D}$). The observed value of the poloidal flux used for the spheromak simulation is the reconnected flux of $1.4\cdot10^{14}$, which is used to derive the toroidal flux of $1.0\cdot10^{14}$~Wb which makes the total spheromak flux ($\phi_{spheromak}$) $4.8$ times $\phi_{FRi3D}$. We note that this in-situ data fitting technique adds error to the input parameter which will translate into errors in the EUHFORIA simulations. {\color{black}{In principle, the in-situ tool could be applied to observations from spacecraft encountered by the CME closer to (but $>0.1$~au from) the Sun. However, such encounters are rare. Additionally, for the real-time forecast, the procedure of acquiring observations, processing them, using them to constrain the CME parameters, and running the EUHFORIA simulation would last too long. Therefore, this method does not facilitate constraints for performing actual forecasts.}} 

\begin{figure*}[htb!]
    \centering
    \subfloat[FRi3D: Radial velocity]{{\includegraphics[width=0.49\textwidth]{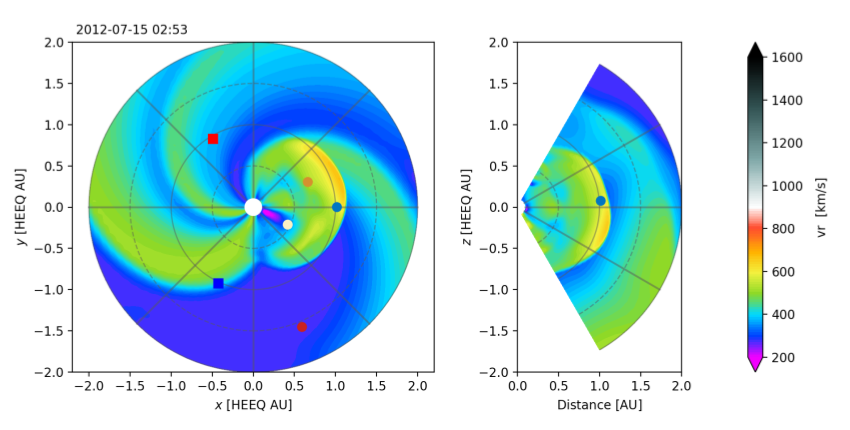} }}
    \subfloat[Spheromak: Radial velocity]{{\includegraphics[width=0.49\textwidth]{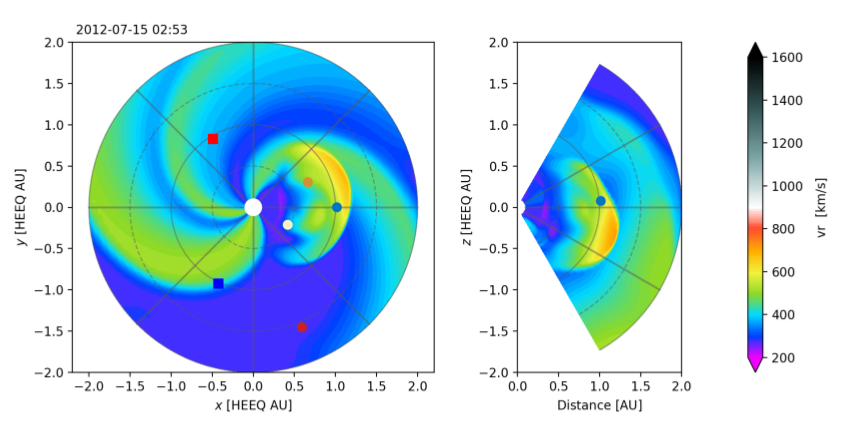} }}\\
    \subfloat[FRi3D: Scaled number density]{{\includegraphics[width=0.49\textwidth]{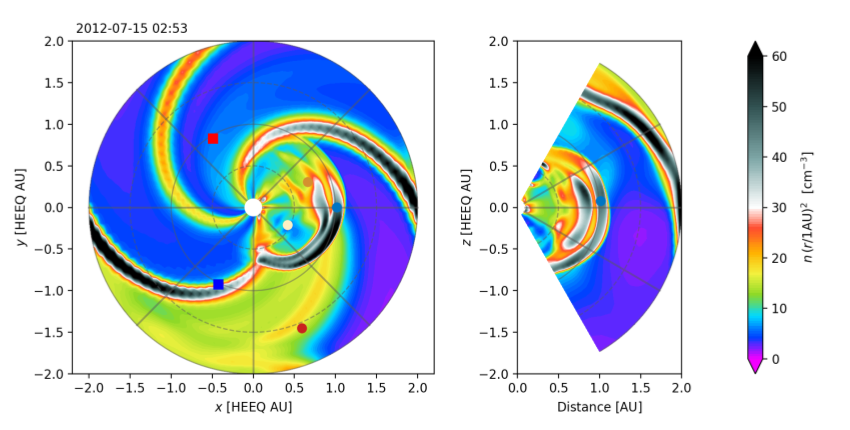} }}
    \subfloat[Spheromak: Scaled number density]{{\includegraphics[width=0.49\textwidth]{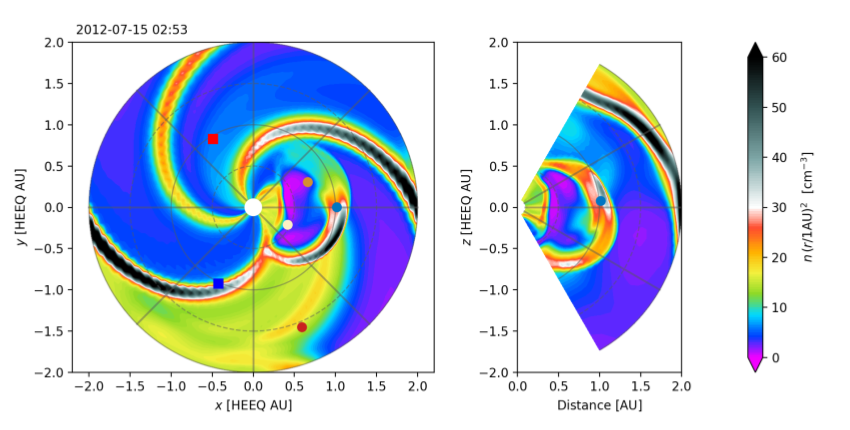} }}\\
    \subfloat[FRi3D: Co-latitudinal magnetic field component]{{\includegraphics[width=0.49\textwidth]{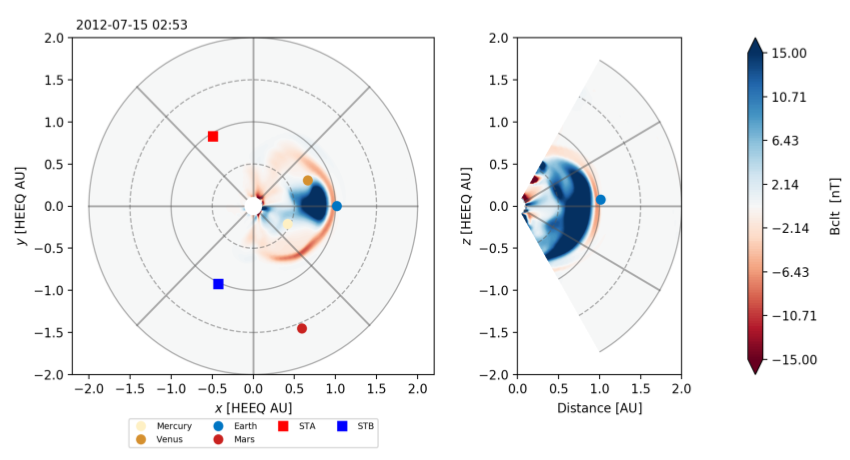} }}
    \subfloat[Spheromak: Co-latitudinal magnetic field component]{{\includegraphics[width=0.49\textwidth]{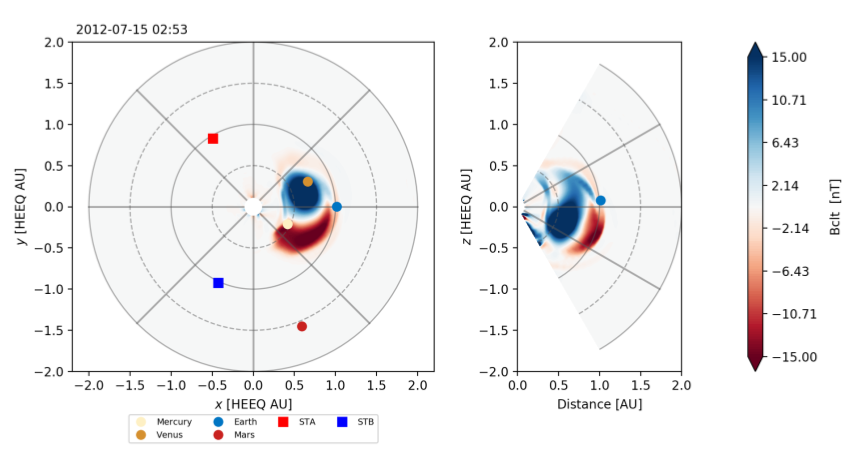} }}
    \caption{EUHFORIA snapshots of the CME event that erupted from the Sun on July 12, 2012. \textit{Top to bottom}: radial velocity ($v_r [\text{km s}^{-1}]$), scaled number density ($n\cdot(r/1~au)^2 [\text{cm}^{-3}]$) and co-latitudinal magnetic field component ($B_{clt} = -B_z [\text{nT}]$) in the equatorial and meridional planes of EUHFORIA's heliospheric domain are plotted for the FRi3D (left panel, (a), (c) and (e)) and the spheromak simulation (right panel, (b), (d) and (f)) of the event as detailed in Section~\ref{sec:Case_real}. The prolonged negative $B_z$ part of CME can be distinctly observed in the meridional plots of FRi3D.}
    \label{fig:20120712_eq_mer_plot}
\end{figure*}

\begin{figure*}[htb!]
    \centering
    \includegraphics[width=7cm]{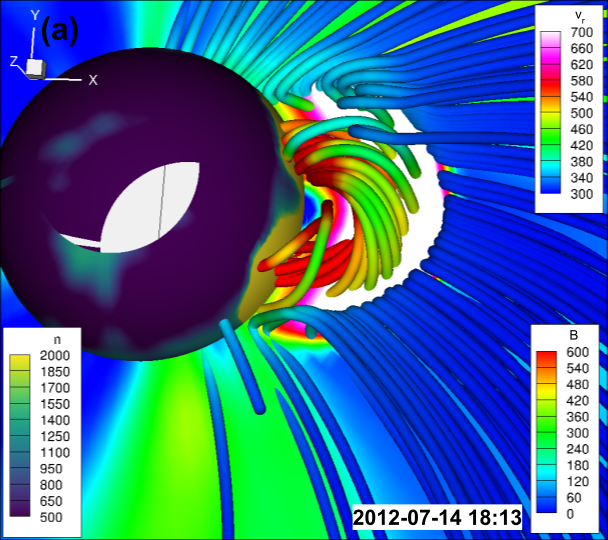} 
    \qquad
    \includegraphics[width=7cm]{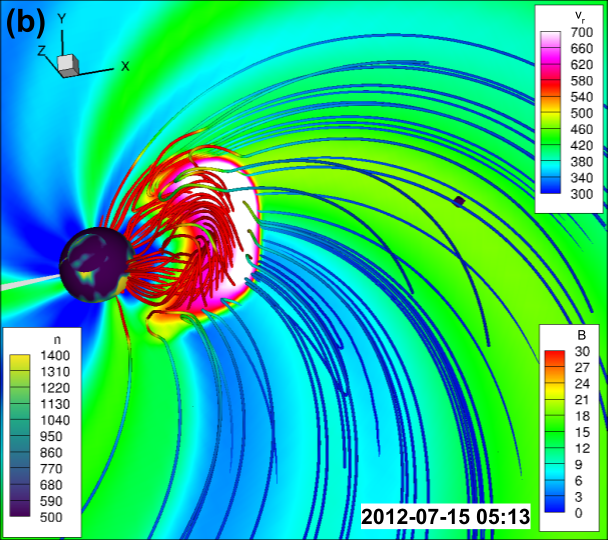} 
    \qquad
    \includegraphics[width=7cm]{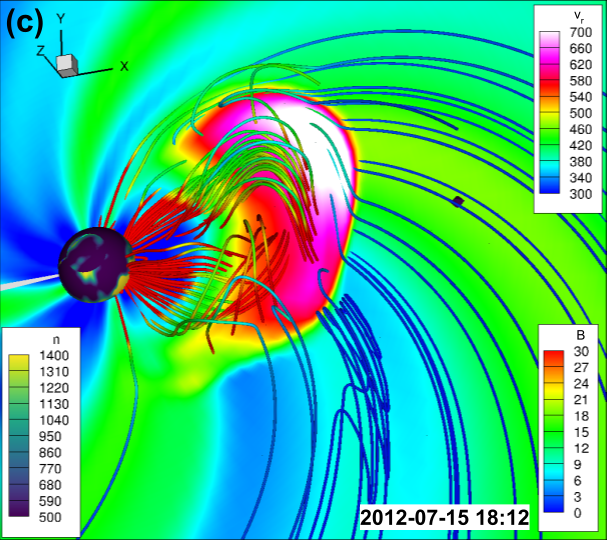} 
    \qquad
    \includegraphics[width=7cm]{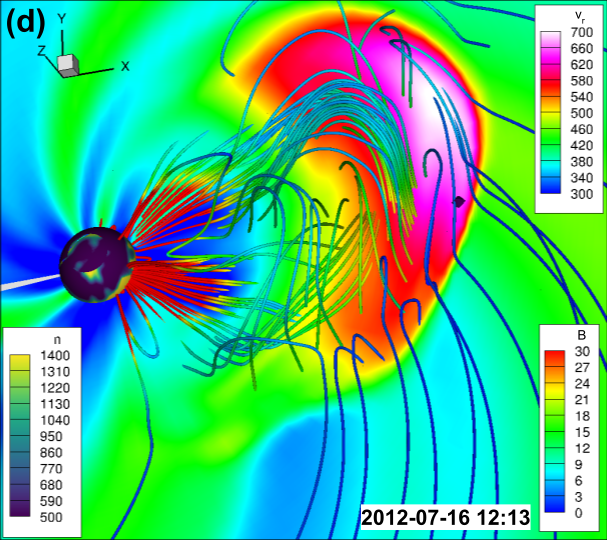} 
    \caption{3D visualization of the EUHFORIA simulation results of the CME that erupted on 12 July 2012 using the FRi3D model, evolving in the heliospheric domain of EUHFORIA. Three colour bars are shown simultaneously: (1) $n$: number density [cm$^{-3}$] on the inner boundary, (2) $v_r$: radial velocity plotted on the equatorial plane, and (3) $B$: total magnetic field plotted on the magnetic field lines. Panel (a) is the zoomed view of the CME emerging out of the inner boundary (the polar region of the inner boundary sphere is excluded \ie, the latitudinal range of the EUHFORIA inner boundary is considered to be $\pm 60^\circ$). {\color{black}{The white region visible through the inner boundary sphere is the space below the equatorial plane where no physical quantity is plotted.}} The strong magnetic field of the CME flux rope can be distinguished from the weaker IMF. At this early phase of evolution close to the inner boundary, minimum interaction between the CME and the background solar wind can be observed. Panels (b), (c), and (d) show snapshots of the evolution of the CME in the heliosphere. The small blue dot on the right portion of these three plots depicts the Earth. } 
    \label{fig:20120712_3D_field}
\end{figure*}

\begin{figure*}[htb!]
    \centering
    \includegraphics[trim={0.5cm 0.3cm 0 0.3cm},clip,width=19cm]{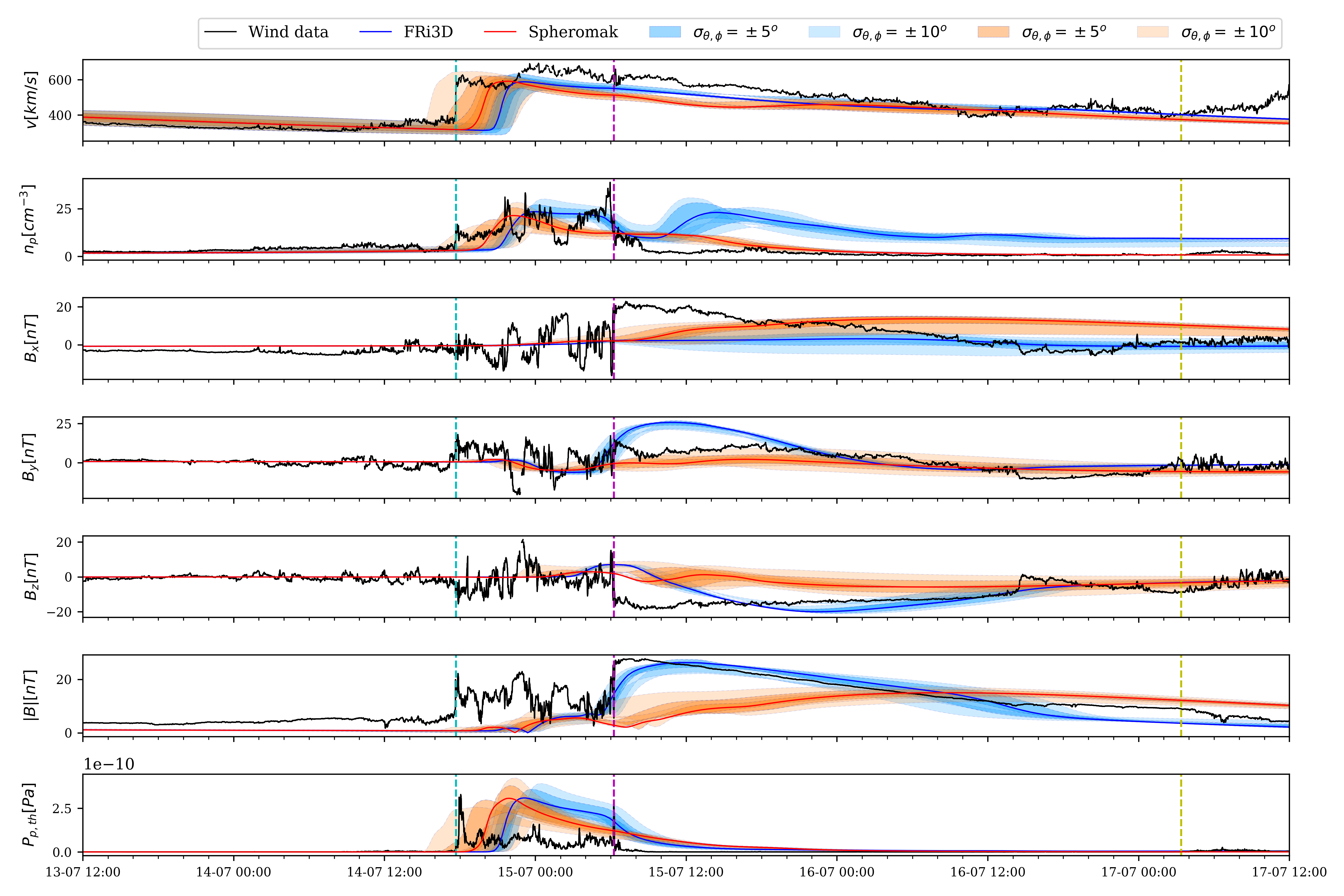}
    \caption{Solar wind properties at Earth upon arrival of CME that emerged from Sun on 12 July 2012. \textit{Top to bottom:} total speed ($v$), proton density ($n_p$), magnetic field components $B_x$, $B_y$, $B_z$ in GSE coordinates, magnetic field strength ($B$) and thermal proton pressure ($P_{p,th}$). Blue solid line time series is obtained with FRi3D CME. The dark and light shade of blue cover the variation of the time series covered by the virtual satellites between $\pm5\degree$ and $\pm10\degree$ of latitude and longitude around Earth. Red and its shades correspond to results obtained from the spheromak CME. Simulation results are compared to the 1-minute average in-situ solar wind measurements of the corresponding properties from Wind in black. {\color{black}{The ICME arrival time, the flux start and end time as recorded in \href{https://wind.nasa.gov/ICME_catalog/ICME_catalog_viewer.php}{Wind ICME catalog} are marked with cyan, magenta, and yellow lines respectively.}}}
    \label{fig:20120712}
\end{figure*}

The CMEs are initialized with a uniform temperature of $0.8$~MK, which is the standard value used in EUHFORIA simulations \citep[][]{Pomoell2018}. Following the results of \citet[][]{Temmer2021}, the mass density is set to $10^{-17}$~kg~m$^{-3}$ inside the FRi3D CME. For the spheromak CME, we choose the same density as in \citet[][]{Scolini2019}, \ie, $10^{-18}$~kg~m$^{-3}$. We obtained the initial momentum of the CMEs launched using the FRi3D and the spheromak model to be
$7.9\cdot10^{18}$~kg~m~s$^{-1}$ and $5.1\cdot10^{18}$~kg~m~s$^{-1}$, respectively. This result was found employing the mass $1.2\cdot 10^{13}$~kg and $0.8\cdot 10^{13}$~kg, and speed $664$~km~s$^{-1}$ and $763$~km~s$^{-1}$ for FRi3D and the spheromak respectively. 
Both models exhibit different expansion rates due to their different magnetic field configurations and this should be also taken into account while determining the injection speed. Therefore, FRi3D CME must be launched with the toroidal speed which can be significantly different from the radial speed estimated from GCS fitting. {\color{black}{The complete list of the input parameter for EUHFORIA simulations with the spheromak and FRi3D models for this case study can be found in Table~\ref{tab:euh_params}.}}

 


\subsection*{EUHFORIA setup}
The EUHFORIA simulations are performed using a spherical grid with a resolution of $1.6 R_\odot$ in the radial direction, $4^\circ$ in the latitudinal direction, and $2^\circ$ in the longitudinal direction. The simulation results are illustrated in Figure~\ref{fig:20120712_eq_mer_plot}. We compare radial velocity, scaled density, and co-latitudinal magnetic field component ($B_{\rm clt}$ = $-B_z$) profiles at a particular instant in the simulation domain, using the spheromak and FRi3D models. The {\color{black}{wider}} longitudinal extent of the FRi3D flux rope {\color{black}{compared to the spheromak}} can be seen in the equatorial cross-sections {\color{black}{in Figure~\ref{fig:20120712_eq_mer_plot}(a), (c) and (e)}}. The evolution of the FRi3D flux rope is illustrated in Figure~\ref{fig:20120712_3D_field}. The CME field lines can be distinguished from the background solar wind magnetic field. With the expansion of the evolving CME, the internal magnetic field can be seen to be decreasing. 

\subsection{Comparison of EUHFORIA results with observations}

Results of the EUHFORIA simulations employing the spheromak and FRi3D CMEs are shown in Figure~\ref{fig:20120712}, compared to 1-minute average in-situ measurements of solar wind properties from \href{https://omniweb.gsfc.nasa.gov/ftpbrowser/wind_min_merge.html}{Wind} in GSE coordinates. In this section, we discuss the simulated ICME arrival time, peak speed, and magnetic field components, focusing on quantifying the improvement of the magnetic field profile using FRi3D over the spheromak. The time series at Earth are plotted using solid lines - the spheromak in red and FRi3D in blue. The variability of plasma parameters ($\sigma_{\theta,\phi}$) in the vicinity of Earth is captured by placing virtual spacecraft at $\pm$ 5$\degree$ and $\pm$ 10$\degree$ in latitude ($\theta$) and longitude ($\phi$), and is indicated by the shaded regions for each CME model around its time series at Earth. 
The speed profiles in panel~1, show the CME-driven shock arrival time of the spheromak and FRi3D. Both profiles are delayed with respect to the observed arrival time of the CME shock. The delay is  $\sim$ 1 hour and $\sim$ 3 hours for the spheromak and FRi3D, respectively. 
The difference in the arrival time at virtual spacecraft (in the area around Earth), can be used to quantify the uncertainty in the CME launch parameters using the initial forward modelling reconstruction.
While the predicted arrival at $\pm$10$\degree$ varies up to $\pm$ 5 hours with respect to Earth for the spheromak, the uncertainty spans $\pm$ 1 hour for FRi3D. Both CME models impact Earth with a maximum speed corresponding to $\sim$ 86\% of the observed maximum speed. 
Bench-marking our CME shock arrival time predictions, we find a good agreement  of our arrival uncertainties with the prediction delays by best-performing CME models \ie, average arrival time within $\pm10$ hours compared to observations with a standard deviation of 20 hours \citep[][]{Riley2018(b),Mays2015}. 

The observed maximum $B$ during the ICME passage corresponds to $27$~nT. While {\color{black}{the}} spheromak simulation predicts $15$~nT (corresponding to $\sim55\%$ of the observed maximum $B$), FRi3D predicts $25$~nT (corresponding to $\sim92\%$ of observed maximum B). Hence, FRi3D achieves a relative improvement of $\sim37\%$ with respect to the spheromak for the prediction of the maximum magnetic field strength upon impact. 
The $B_z$ component has a pivotal role in determining the geoeffectiveness of a CME. 
The minimum $B_z$ values, predicted by the spheromak and FRi3D CME models, are about $-6$~nT and $-19$~nT, respectively. These values correspond to $\sim33\%$ and $\sim105\%$ of the observed minimum $B_z$. 
FRi3D misses the $B_x$ peak but captures the $B_y$ and $B_z$ components, adding up to an almost accurate fit of the total magnetic field. 
The extended flux rope geometry clearly performs better in reproducing the prolonged southward $B_z$ however, the minimum $B_z$ dip is delayed by almost 12 hours. {\color{black}{The observed $B_z$, {\color{black}{albeit reaching}} the minimum value earlier, stays almost constant {\color{black}{until}} around the time when the $B_z$ simulated by FRi3D dips. Qualitatively, the extended profile has been reproduced. We plan to do more event studies in the future, in order to better address this issue}}. The observed elongated $B_z$ profile of the ICME is a signature of the part of the CME magnetic cloud between the apex and the flank crossing the spacecraft \citep[see e.g., Figs. $8$ and $9$ in ][]{Marubashi2007}. 
This is in accordance with the predicted impact based on the 3D reconstruction, which suggested that the CME was launched slightly obliquely, below the equatorial plane, east of the Sun-Earth line, and tilted by $45\degree$. 
The larger longitudinal extent due to the FRi3D is able to capture the prolonged $B_z$ {\color{black}{as a result of the CME crossing the Earth through a region between the CME apex and the flank}}. This feature is missed by the spheromak due to its compact spherical shape.
The thermal pressure is modelled similarly with both models. Further investigation has to be performed to better model the density profile of FRi3D. 
\subsection{Geoeffectiveness predictions}
\begin{figure*}[htb!]
    \centering
    \includegraphics[trim={0.75cm 1.8cm 0 2cm},clip,width=20cm,height=17cm]{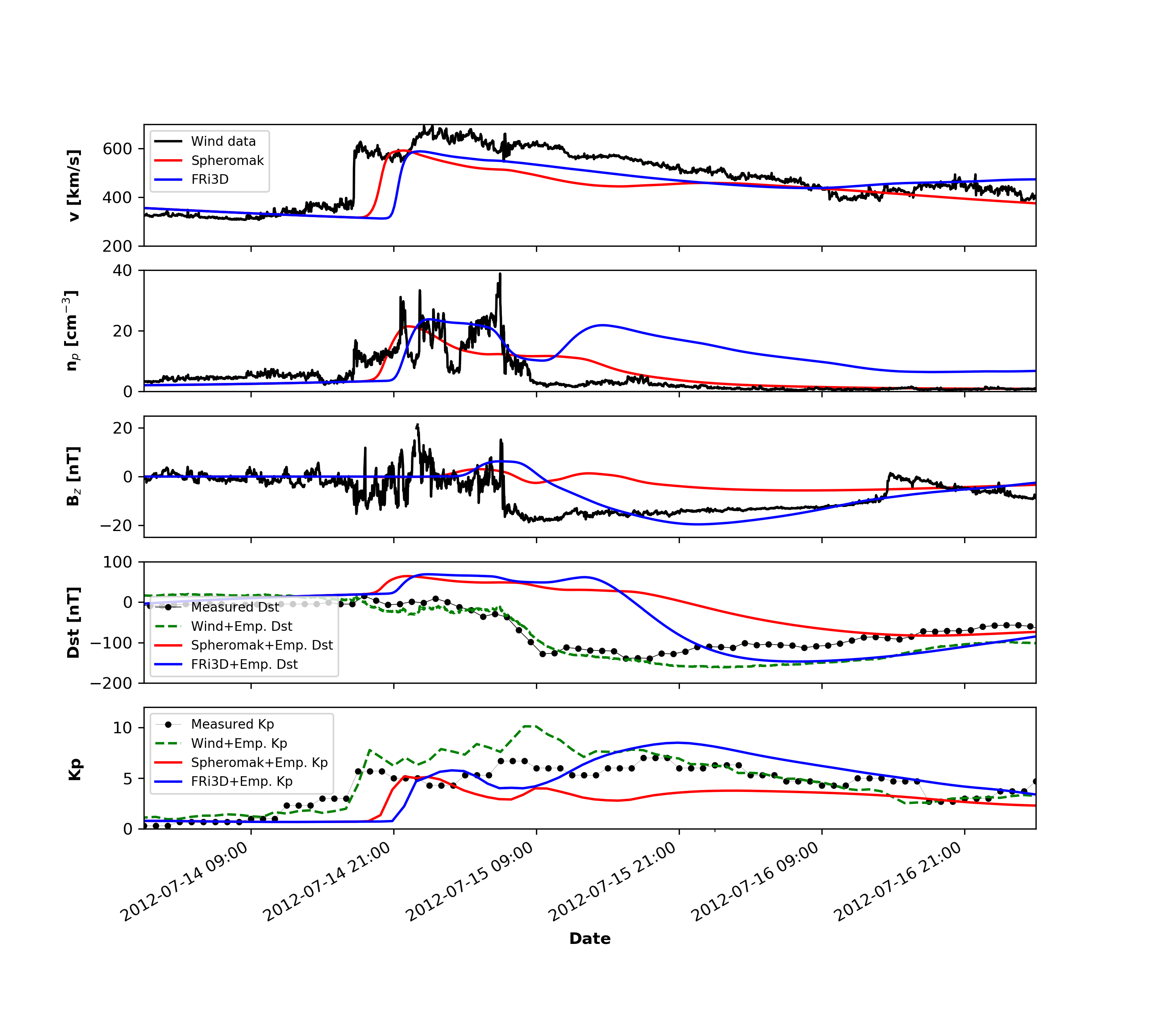} 
    \caption{Comparison of the geoeffectiveness predictions employing the empirical $Dst$ formalism of \citet[][]{Obrien2000a,Obrien2000b} and the empirical Kp-index formalism of \citet[][]{Newell2007,Newell2008} with observations. The empirical $Dst$ (panel 4) and Kp index (panel 5) computed using the measured {\color{black}{Wind}} data (green dashed line), and EUHFORIA simulated solar wind data using the spheromak (in blue solid line) and FRi3D (in red solid line) are compared with their measured values. The solar wind parameters ($v$, $n_p$, $B_z$) at Earth are additionally plotted to show their correlation with the geomagnetic indices. For example, $B_z$ and $n_p$ strongly influence $Dst$ and Kp index respectively.}
    \label{fig:20120712dst_kp}
\end{figure*}

{\color{black}{We employ empirical models to provide early predictions of geomagnetic indices using the simulated solar wind properties obtained from the space weather forecasting tools like EUHFORIA before the CME reaches L1.}} To quantify the CME's geomagnetic impact predicted by EUHFORIA, we use the plasma and magnetic field data obtained by the spheromak and FRi3D simulations to compute the following empirical geomagnetic indices: 
\begin{enumerate}
    \item $Dst$: This is computed using the AK2 model of \citet{Obrien2000a}:
\begin{equation}
    \frac{d}{dt} Dst^* =Q(t) - \frac{Dst^*}{\tau},
\end{equation}
where $Dst^*$ is the corrected $Dst$ after removal of the contamination by magnetopause currents, $Q(t)$ is the rate of energy injection into the ring current and $\tau$ is the decay time. $Q(t)$ is defined as:
\begin{equation}
    Q(t) = -4.4(VBs - 0.5)
\end{equation} 
{\color{black}{where $V$ is the solar wind speed, $Bs = 0$, if $B_z>0$ and $Bs = B_z$, if $B_z<0$.}} The correction is introduced through the effect of the dynamic pressure $P_{dyn}$ \citep[see Eq.~3 in][]{Obrien2000b}. The final $Dst$ (that is compared with observations) is obtained as follows:
\begin{equation}
    Dst = Dst^* + b \sqrt{P_{dyn}} - c,
\end{equation}
where $b$ and $c$ are listed in Table~1 in \citet{Obrien2000b}.
{\color{black}{We compute the empirical Dst using both the observed solar wind data and the EUHFORIA data at Earth location, simulated using the spheromak and FRi3D CME models. Empirical models may have errors associated with their predictions. Therefore, to ensure a consistent comparison we consider it more appropriate to compare results obtained using EUHFORIA time series as input, with results obtained using actual data as input (which provide the best possible prediction from a given empirical model), than with Dst values measured on the ground.}} In panel~4 of Figure~\ref{fig:20120712dst_kp}, the green dashed line shows the empirical $Dst$, computed using the {\color{black}{Wind}} measurements of the solar wind properties at Earth. This estimate of $Dst$ can be used as a reference to quantify the performance of the empirical $Dst$ computed using the EUHFORIA simulations (FRi3D in blue and spheromak in red). The empirical Dst profiles are compared to the observed Dst from \href{http://wdc.kugi.kyoto-u.ac.jp/dstdir/index.html}{WDC for Geomagnetism, Kyoto}. The observed $Dst$ for this event is $-139$~nT. The minimum empirical Dst predictions using FRi3D, the spheromak, and Wind data are $-147$~nT, $-119$~nT, and $-161$~nT respectively. The model overestimates the $Dst$ even with the Wind data. We first compare the improvement in prediction by using FRi3D over the spheromak, with respect to the measured data. Having improved the $B_z$ strength using FRi3D, the minimum $Dst$ is $\sim 20\%$ better than the spheromak simulation [$100\times(min(Dst_{FRi3D}) - min(Dst_{spheromak}))/min(Dst_{observed})$]. Next, we compute the improvement in the performance of FRi3D over the spheromak, with respect to the reference model i.e., empirical $Dst$ computed with Wind data. The minimum $Dst$ predicted by FRi3D is $\sim 17\%$ better than the spheromak with respect to the reference set by computing empirical $Dst$ using {\color{black}{Wind}} measurements [$100\times(min(Dst_{FRi3D}) - min(Dst_{spheromak}))/min(Dst_{Wind\_ref})$]. 
The delay in the $Dst$ dip in the spheromak and FRi3D simulations can be attributed to the {\color{black}{absence of a clearly defined sheath region and the fact that $B_z$ peaks later in the simulations.}} 

    \item Kp index: This proxy for CME geoeffectiveness is calculated in terms of a solar wind coupling function with the magnetosphere \citep{Newell2008} and is given by:
\begin{equation}
    K_p = 0.05 + 2.244\cdot10^{-4} \frac{d}{dt}\Phi_{MP} + 2.844\cdot10^{-6}n^{1/2}v^2.
\end{equation}
The quantity $\frac{d}{dt}\Phi_{MP}$ is the coupling function expressed as the rate of magnetic flux entering the magnetopause and is given by \citep[][]{Newell2007}
\begin{equation}
    \frac{d}{dt}\Phi_{MP} = v^{4/3}B^{2/3}\sin^{8/3}(\theta_c/2),
\end{equation}
where $v$~[km~s$^{-1}$], $n$~[cm$^{-3}$] and $B$~[nT] are the magnitude of speed, number density and magnetic field; $\theta_c = \arctan(B_y/B_z)$. The Kp index mainly depends on the number density and the speed during the CME impact. The maximum observed Kp index (\href{https://www.gfz-potsdam.de/en/kp-index/}{German Research Centre for Geosciences (GFZ)}) for this event is 7. The maximum empirical Kp index obtained using FRi3D, the spheromak and Wind data are 8, 5 and 10 respectively. The empirical Kp is computed using solar wind measurements from {\color{black}{Wind}} and the EUHFORIA simulations{\color{black}{, and compared with observed Kp in the panel~5 of Figure~\ref{fig:20120712dst_kp}.}} 
The maximum Kp predicted by FRi3D is $\sim48\%$ higher than the one predicted by the spheromak [$100\times(max(Kp_{FRi3D}) - max(Kp_{spheromak}))/max(Kp_{measured})$]. 
With respect to the predictions of the empirical model using measured {\color{black}{Wind}} data, FRi3D predictions of the maximum Kp are improved by $\sim33\%$ over the spheromak predictions [$100\times(max(Kp_{FRi3D}) - max(Kp_{spheromak}))/max(Kp_{Wind\_ref})$]. 
\end{enumerate}

\section{Summary and conclusion}
\label{sec:Discussion}
In this paper, we presented the implementation of the magnetized FRi3D CME model in EUHFORIA. The model was {\color{black}{demonstrated by simulating}} synthetic and observed CME events. 
We have performed a synthetic event study (Section \ref{sec:Case_synthetic}) in order to be able to compare {\color{black}{the existing spheromak model}} with FRi3D, and outline the advantages of the new CME model. In Section\ref{sec:Case_real}, we employed the FRi3D model to simulate the observed CME event of 12 July 2012. This event was previously studied with EUHFORIA using the spheromak model \citep[][]{Scolini2019}. We discuss the input parameters of the new CME model and the possible ways to obtain them from observations using the tools included in the FRi3D package. 

The procedure of obtaining the kinematic input parameters {\color{black}{for FRi3D model}} is similar to GCS and takes a similar effort for constraining parameters for the spheromak model. We use EUV observations to constrain some of the magnetic field parameters like chirality and polarity to reduce the number of free parameters in the in-situ data fitting algorithm of FRi3D. The geometrical parameters obtained using the FRi3D forward modelling tool correspond well to those obtained from the GCS tool. {\color{black}{In addition, the magnetic field parameters are constrained using the FRi3D in-situ fitting algorithm. In particular, the constrained magnetic flux is lower than that obtained using remote-sensing observational techniques to constrain the flux for the spheromak simulations. Moreover, the simulation obtained using this flux value is a better fit for the observations. Although constraining parameters using in-situ data can be useful for studies of historic events, this technique is not applicable for space weather forecasting purposes.}}

The main findings of this study as listed below:
\begin{itemize}
    \item The analysis of the synthetic event showed that it is better to employ a higher CME mass density \citep[][]{Temmer2021} for FRi3D than the conventional EUHFORIA density \citep[][]{Pomoell2018}{\color{black}{, since it results in a more reasonable number density profile at Earth}}. 
    \item The synthetic event study showed the improvements in the speed and the magnetic field predictions at the CME flanks by employing FRi3D over the spheromak model.
    \item In the event study of 12 July 2012, we concluded that the toroidal speed of FRi3D is a function of its half-height and differs from the radial speed constrained using the GCS reconstruction for the spheromak. Therefore, we constrained the FRi3D launch speed  using the FRi3D fitting and not the GCS fitting.
    \item Employing the launch speed computed with FRi3D fitting and the increased density, for the event study (12 July 2012), allowed us to analyze the evolution of an observed CME with the FRi3D model. We found an improvement of the magnetic field strength prediction up to $\sim40\%$; the $Dst$ and Kp index were improved up to $\sim20\%$ and $\sim37\%$ respectively, as compared to the simulation using the spheromak model.
    \item  {\color{black}{We show that FRi3D with its global geometry performs better, for the studied event, in reproducing the magnitude of magnetic field components due to its complex design. However, {\color{black}{a variety of CME events need to be studied in order to understand all the drawbacks of the model and to further improve the modelling of the magnetic field components within the CME with FRi3D}}. 
    \item We also found that FRi3D is computationally more demanding than the spheromak model. While computing the CME mask at the inner boundary, it is necessary to find the location of a given point relative to the FRi3D axis to obtain the modelled magnetic field at that point. {\color{black}{Although it is easy to analytically construct individual magnetic field lines for the whole flux rope based on the FRi3D model parameters, it is computationally expensive to deduce magnetic field parameters in a given location in space since it requires numerically inverting the FRi3D equations.}}}} 
\end{itemize}

   This new setup of the FRi3D model in EUHFORIA significantly improves the kinematic description of the CME evolution including the dynamics resulting from CME's interaction with the solar wind. It opens numerous possibilities for future studies, including 3D fits of heliospheric imager and coronagraph data, ideal for space weather forecasting and prediction of CME's geoeffectiveness. 


The implementation of the spheromak in EUHFORIA \citep[][]{Verbeke2019} was followed by optimizing the CME modelling with the spheromak \citep{Asvestari2021}. Similarly, optimization of the modelling with FRi3D within EUHFORIA will be carried out in future studies. The ongoing development efforts aim to enhance the efficiency and speed of CME mask calculation \ie, determining the FRi3D CME cross-section at the inner boundary as the CME enters the heliospheric domain. All these improvements will make FRi3D in EUHFORIA fit for real-time forecasting. 

\section*{Acknowledgments}
 
This project (EUHFORIA 2.0) has received funding from the European Union’s Horizon 2020 research and innovation programme under grant agreement No 870405.
These results were also obtained in the framework of the projects C14/19/089  (C1 project Internal Funds KU Leuven), G.0D07.19N  (FWO-Vlaanderen), SIDC Data Exploitation (ESA Prodex-12), and Belspo projects BR/165/A2/CCSOM and B2/191/P1/SWiM, and the ESA project ``Heliospheric modelling techniques'' (Contract No.\ 4000133080/20/NL/CRS). C.S.\ acknowledges the NASA Living With a Star Jack Eddy Postdoctoral Fellowship Program, administered by UCAR's Cooperative Programs for the Advancement of Earth System Science (CPAESS) under award no.\ NNX16AK22G.\ N.W.\ acknowledges funding from the Research Foundation - Flanders (FWO -- Vlaanderen, fellowship no.\ 1184319N).
The simulations were carried out at the VSC – Flemish Supercomputer Centre, funded by the Hercules Foundation and the Flemish Government – Department EWI.
The authors are very grateful for the invaluable contributions of Dr.\ Emmanuel Chan\'e.

\bibliographystyle{model5-names}
\biboptions{authoryear}
\bibliography{refs}

\appendix

\section{FRi3D volume calculation}
\label{sec:FRi3D_volume}
Consider the flux rope to be an extended cylinder with a variable cross-section for computing the FRi3D CME volume. The radius of the cross-section varies proportionally to the heliocentric distance with the largest radius in the apex of the structure (at the $\phi = 0\degree$) and tending to zero in the Sun as:
\begin{equation}
    R(\phi) = \frac{R_p}{R_t} r(\phi)
\end{equation}
where $r(\phi) = R_t cos^n(a\phi)$ is the cross-section at a given $\phi$ and $a=(\pi/2)/\phi_{hw}$. 
The volume of a cylinder can be calculated as the cross-sectional area of the cylinder multiplied by the height of the cylinder. 
The length of the cylinder is set to the length L of the axis of the CME shell:
\begin{equation}
    L = \int_{-\phi_{hw}}^{\phi_{hw}} \bigg[r(\phi)^2 + (\frac{dr(\phi)}{d\phi})^2\bigg]^{\frac{1}{2}} \,d\phi
\end{equation}

The volume of the FRi3D flux rope is given as:
\begin{align}
    V_{FRi3D} &= \int_{-\phi_{hw}}^{\phi_{hw}} \pi R^2(\phi)\bigg[r^2 + (\frac{dr}{d\phi})^2\bigg]^{\frac{1}{2}} \,d\phi \\
      &= \int_{-\phi_{hw}}^{\phi_{hw}} \pi R_p^2 cos^{2n}(a\phi) \bigg[r^2 + (\frac{dr}{d\phi})^2\bigg]^{\frac{1}{2}} \,d\phi \\
      &= \pi R_p^2 R_t \int_{-\phi_{hw}}^{\phi_{hw}} cos^{2n}(a\phi) [cos^{2n}(a\phi) \nonumber \\
      &+ n^2 a^2 sin^{2} (a\phi) cos^{2(n-1)}(a\phi)]^{\frac{1}{2}} \,d\phi
      \label{eqn:volume}
\end{align}
In this work, Equation~\ref{eqn:volume} is numerically computed.

\end{document}